# The pursuit of stability *in halide perovskites*: the monovalent cation and *the key for surface and bulk self-healing*

D R Ceratti[1]*, A V Cohen[1], R Tenne[2], Y Rakita[1], L Snarski[1], L Cremonesi[3], I Goldian[4], I Kaplan-Ashiri[4], T Bendikov[4], V Kalchenko[5], M Elbaum[6], M A C Potenza[2], L Kronik[1]*, G Hodes[1]*, D Cahen[1]*

[1]Weizmann Institute of Science, Department of Materials and Interfaces, 7610001, Rehovot, Israel.
[2]Weizmann Institute of Science, Department of Physics of Complex Systems, 7610001, Rehovot, Israel.
[3]Department of Physics and CIMAINA, University of Milan, via Celoria, 16 20133, Milan, Italy.
[4]Weizmann Institute of Science, Department of Chemical Research Support, 7610001, Rehovot, Israel.
[5]Weizmann Institute of Science, Department of Veterinary Resources, 7610001, Rehovot, Israel.
[6]Weizmann Institute of Science, Department of Chemical and Biological Physics, 7610001, Rehovot, Israel.

*davide-raffaele.ceratti@weizmann.ac.il , david.cahen@weizmann.ac.il, gary.hodes@weizmann.ac.il, leeor.kronik@weizmann.ac.il

**ABSTRACT** We find significant differences between degradation and healing at the *surface* or in the *bulk* for each of the different $APbBr_3$ single crystals (A=$CH_3NH_3^+$, methylammonium (MA); $HC(NH_2)_2^+$, formamidinium (FA); and cesium, $Cs^+$). Using 1- and 2-photon microscopy and photobleaching we conclude that kinetics dominate the surface, and thermodynamics the bulk stability. Fluorescence-lifetime imaging microscopy, as well as results from several other methods, relate the (damaged) state of the halide perovskite (HaP) after photobleaching to its modified optical and electronic properties. The A cation type strongly influences both the kinetics and the thermodynamics of recovery and degradation: FA heals best the bulk material with faster self-healing; $Cs^+$ protects the surface best, being the least volatile of the A cations and possibly through O-passivation; MA passivates defects via methylamine from photo-dissociation, which binds to $Pb^{2+}$. DFT simulations not only provide insight into the latter conclusion, but also show the importance and stability of the $Br_3^-$ defect. These results rationalize the use of mixed A-cation materials for optimizing both solar cell stability and overall performance of HaP-based devices, and provide a basis for designing new HaP variants.

## 1. Introduction.

Halide Perovskites (HaPs) continue to have significant impact on the field of solar cells, thanks to their very low cost, ease of production, and competitive efficiencies that are comparable, if not superior, to those of established photovoltaic (PV) technologies.[1–3] In the wake of these PV results, other applications, such as light-emission, radiation detection, and electronics are being explored.[4–7] Despite these large efforts, a critical issue hangs as a sword of Damocles over the entire field: stability. Various approaches for delaying or avoiding device degradation, due to external influences such as heat, light, or chemicals, have been tested.[8] Encapsulation of the solar cell and perovskite passivation through long alkylamine molecules[9], as well as use of 2D variants of HaPs.[10,11] all show promise for avoiding degradation; still, the need for decades of stable operation (at least for use in solar cells) under realistic conditions makes the question of the *intrinsic* stability of the materials themselves crucial. We hereafter use the term "intrinsic stability" as the stability of the material irrespective of external (ambient, light, temperature) influence. This can be tested by stressing the material to different extents (in our case for a given laser pulse time and rate, this will be the laser intensity). Intrinsically unstable (or metastable) materials will degrade already with low intensity stresses.

Degradation pathways have been identified for the thermal and photo-decomposition of *methylammonium (MA)* lead halides using mass spectroscopy,[12] STA-FTIR (simultaneous thermal analysis (STA) and Fourier Transform-Infrared (FT-IR) Spectroscopy)[13]. The pathways that were identified likely cause loss of photo-conversion efficiency and involve the release of methylamine, hydrogen halides, halomethanes, ammonia, and halogens. However, such mechanistic analysis is lacking for *formamidinium (FA)* and *cesium (Cs)* lead halides, which are critical building blocks of the currently most-efficient halide perovskite solar cells.[2,14] An important question then follows: *are mechanisms of decomposition connected to an intrinsic material instability or are there ways to eliminate or minimize them*? For example, mechanisms involving the attack of the material by external chemicals are strongly in-hibited by encapsulation and do not affect the intrinsic material stability. However, if degradation pathways not involving external chemicals are unavoidable, HaPs will be practically unusable for long-term operation For this reason, we focus our study on the perovskites themselves; while studies on complete cells are the most relevant ones for applications, using multi-component samples makes determining the limits of stability of the HaP material, the one irreplaceable element of a perovskite solar cell, very difficult and mostly, impossible.

Here, we address these issues by assessing the fundamental differences between surface (chemically accessible) and bulk (depth of ~ 110 μm, chemically inaccessible) stability in halide perovskites single crystals, focusing on the critical role that the A cations play in HaP stability. We show that there are very significant differences between self-healing of perovskites[15] in the bulk and that at surfaces, with major effects on the perceived stability of the material. Specifically, through confocal and fluorescence lifetime imaging microscopy (FLIM), we find that upon illumination the optoelectronic properties of the material change differently at the surface and in the bulk, including whether and how they revert to the original state after photo-damage. To pinpoint the chemical origin of these variations, we analyze the effect of photodamage at the surface and in the near-surface region. Additionally, we use density functional theory (DFT) simulations to rationalize, at a molecular level, the A cation effect. In particular, we analyze how the self-healing directly controls the defect density in the HaPs. We conclude that the thermodynamically-determined stability and defect density in the bulk are the relevant quantities for solar energy applications. The more pronounced surface degradation is due to kinetic effects (material loss), which can be hindered with tailored encapsulation and judicious electron transporting layer (ETL) and hole transporting layer (HTL) choice, making the system as closed as possible, i.e., limiting exposure to ambient and especially escape of eventual possible products as much as possible. These layers must not solubilize or react with the products of decompositions of the HaPs, so as to make the HaP thin film bulk-like.

The results that we present here constitute a major step beyond what was known from earlier work, ours, i.e., that bromide perovskite can self-heal,[15] because *we now identify and explain the mechanisms involved in the self-healing*, an obvious step towards optimizing self-healing. Our results also go well beyond earlier suggested, deduced or assumed self-healing,[16–20] and pave a way of reaching long-term stable halide perovskite solar cells, rationally, i.e., beyond empirically.

Here we use self-healing to describe the series of chemical processes that undo damage in a material or system, without the involvement of any external factors/effects. We distinguish it from self-repair, which is used also to describe the ability to repair more complex systems, not by net reversal of the mechanisms that led to the damage, but through different actions that can involve other factors/effects, triggered without external intervention, to mend the damage.

In the following we present the results, and discuss them. In the results section we first analyze results of photo-damage and self-healing (detected through photoluminescence variations) on the surface and in the bulk. We then examine if these processes affect other properties of the material such as bandgap and carrier lifetime. We continue by analyzing the chemical origin of the self-healing in the bulk through DFT calculations, and conclude by getting chemical information on the effect of the photobleaching on the surface from different analysis methods (AFM, EDS, XPS). All these results then constitute the basis of the following discussion on the different roles of the A cation for the stability of the perovskite in the bulk and at the surface. We also connect self-healing to the overall stability of the material and, most importantly, to the low defect density, also in steady state *under illumination*. We conclude by stressing the importance of encapsulation, not only to avoid the penetration of chemicals from outside, but, as importantly, to prevent the escape of perovskite decomposition products, which are necessary for self-healing after damage.

**2. Results: surface and bulk self-healing**

*2.1 Time dependence of photoluminescence (PL) intensity.*
We follow the PL of HaPs using an experiment that is, essentially, one of fluorescence recovery after photobleaching (FRAP), which is of common use in biological experiments involving fluorophores (see Supplementary Information (SI.2-3) for a detailed description which is required for a proper calculation of the energy absorbed by the material (SI.4-5)). A FRAP experiment usually comprises three steps: (1) Obtain the PL image before damage (Figure S2-A), to be used as the reference PL signal, using a low laser power (LP). (2) Cause damage (bleaching) using high LP (Figure S2-B); (3) Monitor PL recovery over time, at imaging LP as in step 1 (Figure S2-C-D). We use a variant of the method previously employed for the bulk[15] to assess the (near)surface damage and recovery of the APbBr$_3$ single crystals (see Supplementary Information SI.3 for the procedure). In brief, a laser beam is focused on the sample surface through a microscope objective with low LP, exciting carriers by 1-photon absorption. The excited carriers recombine, inducing PL. The light is collected from the focus and used to obtain a reference image. Beyond a threshold excitation laser power (LP = 50*(imaging LP) = 0.88 mW (488 nm)), damage is observed (i.e., photobleaching). The intense laser beam is focused on the surface and scanned over (8µs per pixel) defined circular regions (~ 8 µm diameter). In this way, the bleached areas are clearly distinguishable from the background in the microscope images. The laser is absorbed to a ~ 200 nm depth, as defined by the 1-photon absorption coefficient of the material[21]. The photo-generated carriers diffuse, releasing the energy in a volume determined by their diffusion length (which can reach values of µm[22]); in the following, surface damage is always related to this "near surface" process. Subsequently, the sample is irradiated again with lower LP for continuous imaging, yielding a time series of images of the bleached area. PL change, associated with sample modification, is observed only where damage occurred. At low LP any damage, if present, is healed in the time between successive images.

Because HaPs are of particular interest for PV cells, we compare the energies used to cause the damage to normal solar illumination. To relate 1-photon bleaching to normal solar illumination, we compare the absorbed energy per unit volume (J·m$^{-3}$). Our 1-photon measurements deliver energies / volume equivalent to 1 - 100 seconds of solar illumination (the detailed calculation of the energy deployed on the sample, requiring proper discussion, is reported in SI.4 and SI.5). Irradiation generates carriers in the sample. These recombine, releasing heat and, in some cases, activating chemical processes. As heat can induce damage to the perovskite, we evaluate the irradiation-induced change in temperature in SI.6. We found variations of temperature $\Delta T < 10°C$, which are small to be relevant for heat-induced damage. We are not aware of methods for experimental evaluation of the temperature increase inside the material. Because of the inaccessibility of the bleaching site, it would require optical detection confocal with the bleaching. Infrared characterization could, for example, be used, but detection would need good temporal resolution (µs) and be extremely sensitive. Thermal radiation is very low at moderate temperature and would originate in the small volume defined by the diffraction limit of the laser (hot-carrier relaxation) or, for a more approximate evaluation, the diffusion length of the carriers in the order of few µm$^3$.

We stress that self-healing takes place whatever the origin of the damage and is relevant on short and long timescales. For the latter it was argued that temperature is critical for long-terms PCE stability[23,24].

Data on the stability and self-healing of the perovskite in the bulk cannot be obtained by focusing the 488 nm laser into the bulk because of the high absorption of the perovskites. It is, however, possible to circumvent this limitation by using 800 nm light and exploiting 2-photon absorption. Here, pulsed 800 nm laser light is focused through the same microscope objective to a plane well inside APbBr$_3$ crystals, which are transparent at this wavelength. In the focus, the high intensity of the laser stimulates the non-linear absorption of the 800 nm light. Carriers are generated and the 2-photon stimulated PL is collected as in the preceding case. The damage and monitoring procedures are performed following the protocol of the 1-photon experiment. The calculation of the absorbed energy per unit volume and the temperature increase due to 2-photon excitation can be found in SI.4-6 and yields values similar to the ones found in the 1-photon case. Because of the non-linear characteristic of 2-photon absorption and the importance of optical effects on the amount of energy absorbed from the material, it is not possible to provide a quick evaluation of the 2P absorbed energy, but this is treated in-depth in SI.4 and SI.5. In Figure 1, we report the background normalized PL just after bleaching (solid lines) and after recovery (dashed lines) as a function of the LP (in % of (1P - 488 nm) 880 µW and (2P - 800 nm) 771 mW). For all samples, the most striking feature is that we detect a sharp, sample-dependent threshold for damage in the 1P-surface experiment, but not in the 2P-bulk one. In the latter, the effects of photo-bleaching are more gradual (note that for 2P or 1P bleaching, the deposited energies grow quadratically or linearly, respectively, with the % of LP).

At the surface, MAPbBr$_3$ crystals appear to be the most resistant to photo-damage: with 1P-surface excitation, the LP needs to reach 60% of its maximal value whereas for FA and Cs 45-50 % suffices. However, once damaged, the MA material heals only slightly (PL recovers 15% of its original value), in contrast to FA and Cs based samples, which show partial to almost complete healing after 12h. For 1-LP > 80% the Cs

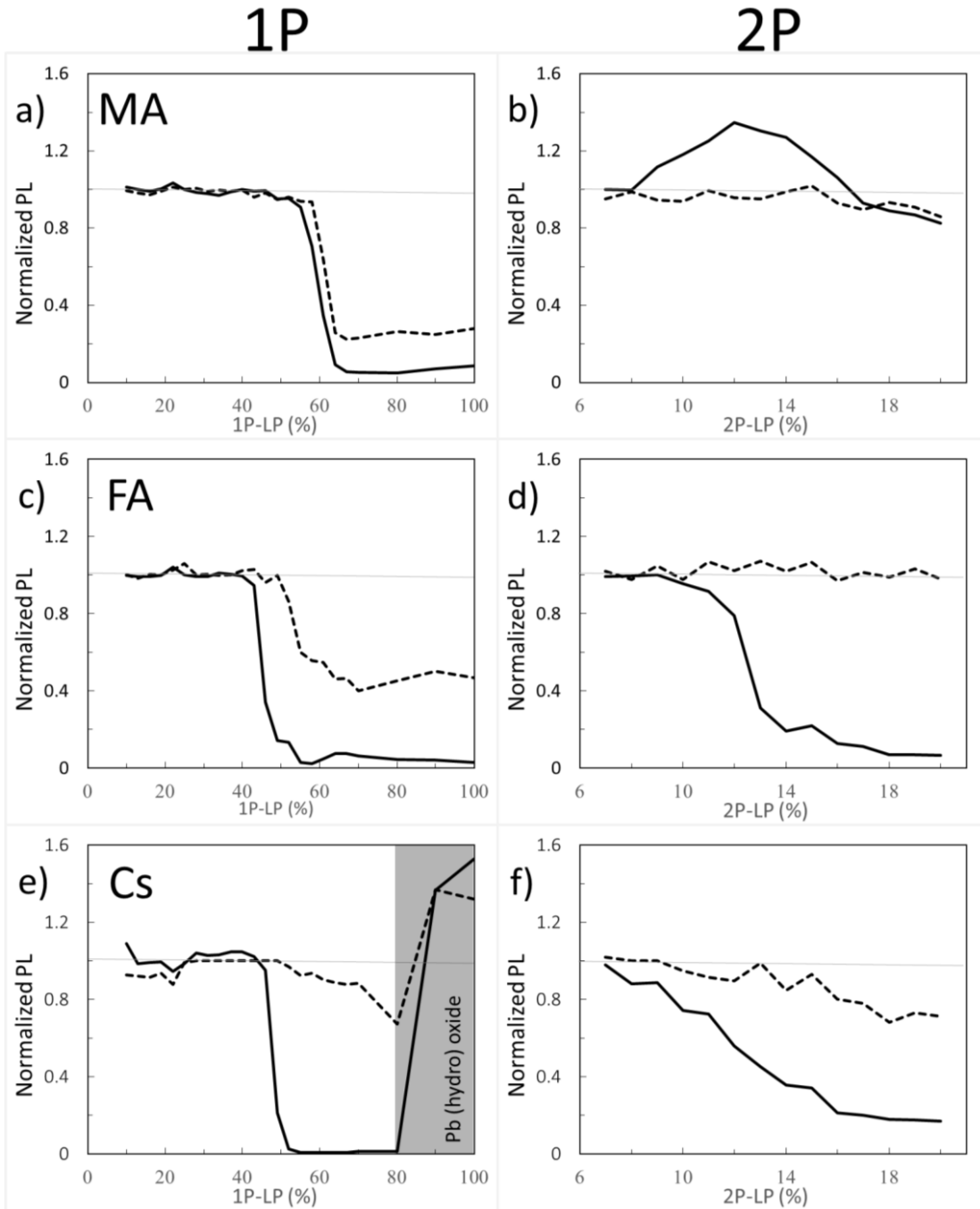

**Figure 1.** Background-normalized PL intensity of (**a, b**) MA, (**c, d**) FA, and (**e, f**) Cs lead tribromide perovskite samples, immediately after bleaching (solid line) and after 12 h of recovery (dashed line) for the 1P (**left**) and 2P (**right**) experiments. The laser power (LP) is expressed in %, with 100 % (1P) being 880 μW and (2P) 771 mW. For Cs, we see in the 1P experiment a large PL increase exceeding the initial level at > 80 % LP as marked by the gray shading. This corresponds to a new, blue-shifted phase discussed in the text.

HaP surface even shows increased PL intensity. Surface behavior can be influenced, usually negatively, by pre-existing surface defects that are known to influence the chemical reactivity and the stability of HaPs.[25,26] Detecting and measuring the effects of surface defects on self-healing, even if not yet achievable today, will constitute a major step in stabilizing polycrystalline thin perovskite films.

Analyzing the damage by 2P-bulk excitation, we see that the increase in damage always gradually increases with 2P-LP. Self-healing is more effective with an organic A cation (Figure 1.b, 1.d) than with Cs (Figure 1.f). In particular, MA is subject to two processes after photo-bleaching: (1) for low to medium LP (with energies equivalent to 30-50 seconds of normal sun irradiation) the process increases the PL; (2) for high LP (50-80 seconds of sun irradiation) the process decreases PL intensity. After bleaching, the system reverts to the original state for both processes. Process (1) requires recovery times of the order of 30 sec while process (2) requires several minutes. FA and Cs samples show a single process similar to process (2) of MA. After the bleaching, FA heals completely and very rapidly (< 30 sec).

For low 2P-LP the healing is so rapid that the PL is already partially recovered within the time that elapses between the end of the 2P excitation and the PL measurement of the first data-point (5 seconds). Cs HaP, however, is healed on much longer timescales (> 12 h).

To conclude this section we stress that, even though our results are valid for a mechanism that takes place after a short pulse of irradiation, as the mechanism is always present in the material, it is critical also for long-term operation, equivalent to continuous damage and healing processes! As long as the time scale of healing is short enough to prevent damage accumulation, self-healing can influence the material properties also for extremely long times. Examples are Si:Li, used in commercial radiation detectors, and CIGS, used in commercial solar cells.

*2.2 PL spectroscopy.* PL spectroscopy can indicate modification of material composition and/or structure that is damage/healing-induced, beyond changes in PL intensity. Figure 2a shows PL emission spectra recorded for the three samples after bleaching through 2P excitation (for experimental settings see SI.3). Bleaching was performed at the surface, with the exception of MA with medium intensity that was bleached slightly inside (~ 2.5 μm) the crystal. The spectra were recorded using 2P excitation at/near the surface because of three reasons: (1) using 2P excitation, all collected PL reaches the detector, thus avoiding any filtering which would be necessary to eliminate the 1P exciting light. The collected spectra are then more reliable than for a 1P measurement; (2) 2P excitation in the (semi)bulk is the only way to modify MA to its transient, more PL intense, state. Note that focusing the 2P light precisely at the surface requires extreme attention as slight inclination of the sample can shift the behavior from surface-like to bulk-like; (3) spectra from the bulk (~110 μm) are modified by self-absorption effects. Photons emitted from the bulk pass through the material, which absorbs more efficiently the higher energy photons. As a result, the PL peak from the bulk is blue-filtered and any true blue-shift due to chemical or structural variations is masked. Focusing on the surface with a 2-photon excitation, we excite the carriers in a volume that reaches deeper into the material (~ 1.5 μm) than what one would obtain with 1P excitation (~ 0.2 μm). As we discussed earlier, the excited carriers diffuse, therefore a slight larger volume emits PL after 2P excitation on the surface. As the diffusion length of the carriers is larger than the focus of the objective through which PL is collected, this effect should not modify the recorded spectra significantly.

Analyzing Figure 2a, we find that neither the more luminescent nor the less luminescent form of MAPbBr$_3$ show a detectable PL peak shift (Figure 2a-MA). Similarly, there is no significant shift of the FAPbBr$_3$ PL peak (note that absolute PL values are reported in Figure 2). In the same way, CsPbBr$_3$ shows no change in the PL emission spectrum as a result of *reversible* bleaching ('medium' in the figure). However, it does show a 6 nm blue shift for areas which exhibit increased PL intensity after bleaching with the highest LP (Figure 2a, Cs-High). This can be attributed either to the formation of lead (hydro)oxide[25,26] or carbonate[27,28] or to the oxygen-related sequestration of Pb$^{2+}$ from the crystal lattice, leaving a Cs-rich material. In the latter case, the blue shift can arise from a contribution by Cs$_4$PbBr$_6$, known to emit at shorter wavelength (500 nm) than CsPbBr$_3$ (530 nm).[29] A Pb-deficient perovskite composition, which would agree with our PL emission data, has been reported,[30] showing how variations in composition blue-shift the PL peak. For completeness, we report that spectra recorded completely in the bulk do not show any peak shift either for the three materials.

*2.3 PL lifetime imaging (Fluorescence Lifetime Imaging Microscopy, FLIM).* The absence of spectral shifts in the perovskite emission suggests that bleaching can influence the chemistry of a material through thermal or photo-induced reactions possibly forming traps. To explore this, we map the PL lifetime of the perovskite crystals both at the surface and in the bulk. Bleaching-created traps should decrease the lifetime of the photo-generated carriers and decrease PL lifetime. Decreasing the lifetime also decreases the PL intensity. Specifically, in a trap-assisted recombination regime (indicated by exponential decay of the instantaneous PL intensity over time), the integrated PL intensity between pulses decreases if the lifetime decreases.

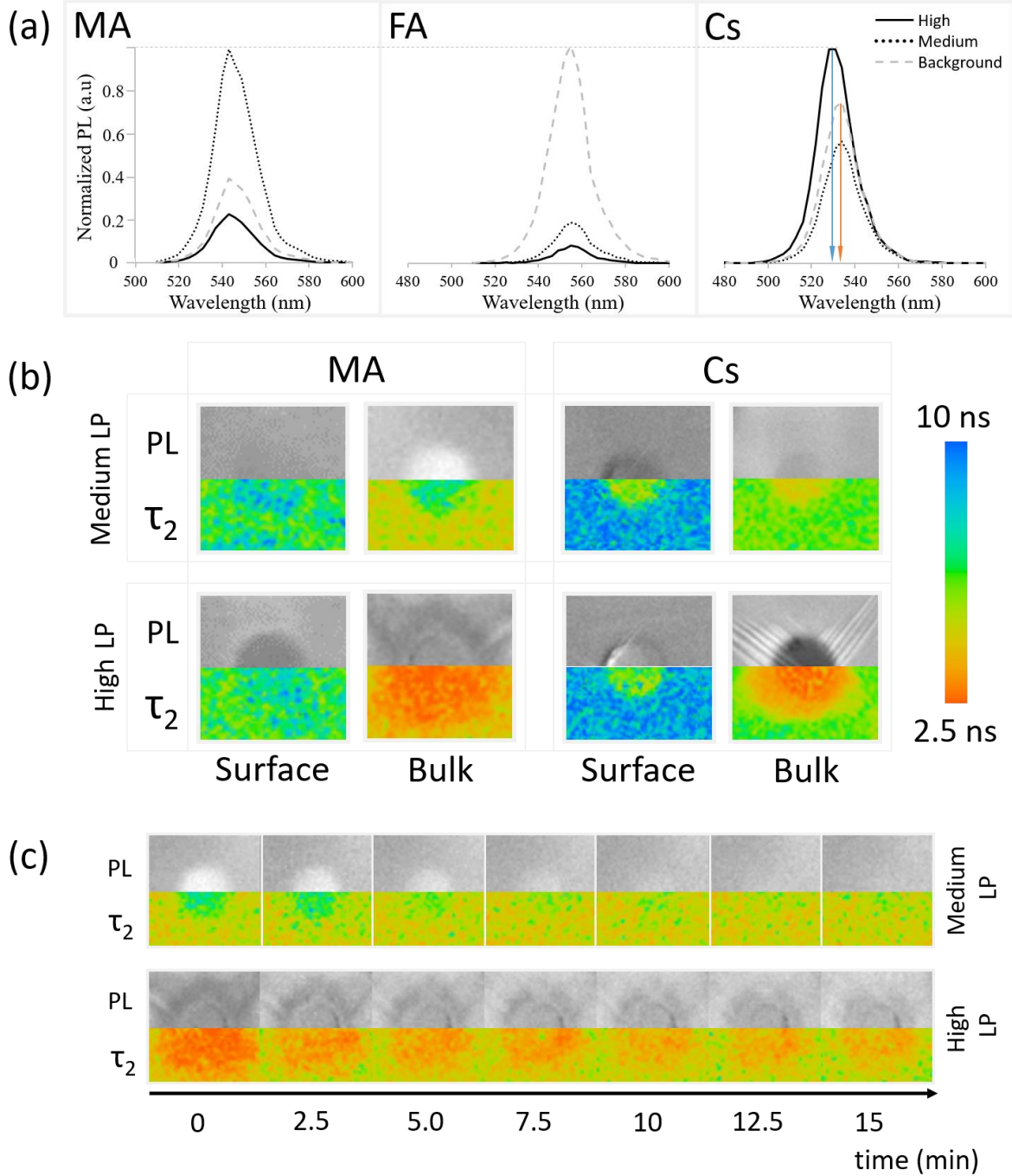

**Figure 2. (a)** Normalized PL spectra from the surfaces of MA-, FA- and Cs-PbBr$_3$. The peaks with the value of 1 are the ones which are most luminescent. The spectra obtained in the other conditions are normalized versus the most luminescent one. Solid line: strongest bleaching. Dotted line: medium bleach. Gray dashed line: background (at 20 μm lateral distance from the bleached area). The arrows in the Cs samples stress the blue shift of the PL peak for the high LP bleach (blue arrow) with respect to that of the background or the medium LP bleach (red arrow). The spectra are normalized to the maximum PL value of each experiment. Medium bleach in the MA case is recorded from slightly (1μm) inside the single crystal. In all the other cases it is recorded on the surface. **(b)** Maps of the PL intensities (grayscale) and maps of PL lifetimes (color-coded – legend on the right side) for the MA and Cs samples on the surface and in the bulk, after medium and strong bleaching. **(c)** Time evolution of the PL (grayscale) and PL lifetime (same color code as in **(b)**) after the bleaching, for the MA sample *in the bulk* at medium LP (Top) and high LP (bottom). Each step corresponds to a delay of 2.5 minutes.

In Figure 2b we relate the PL intensities (top) and FLIM (bottom) measurements of the MA and Cs samples after medium LP (30 seconds of solar illumination) and high LP (100-200 seconds of solar illumination) bleaching on the surface and in the bulk (110 μm). Data for the FA samples are not reported because the lifetime of the photogenerated carriers is of the order of the time delay between the laser pulses (12.5 ns), or longer, which does not allow a proper fit of the decay.

The PL decay is found to be bi-exponential (details of the fitting are given in SI.7). Following SI.7, the fastest decay is due to carrier trapping and the slowest to detrapping. We focus on the slowest decay because it is the kinetically limiting process. Additional information on the shorter lifetime result /processes is reported in SI.7.

Figure 2b-MA-Surface shows that the PL intensity decreases (darker semi-circle, top half) at the surface for both medium and high LP. This decrease is not accompanied by any clear variation in the PL lifetime between the exposed (center of the lifetime image) and non-exposed (side of the lifetime image) volumes. We interpret these observations as a "shadowing" effect due to a non-emitting material formed on the surface. This material blocks the PL originating from the MAPbBr$_3$ lying below it. In addition, the microscope assesses the material up to a depth of ~1 μm, even if focused on the surface. If complete degradation happens at or very close to the surface (within ~100 nm), the overall PL also decreases because the amount of emitting material decreases. To conclude, since the degraded material does not emit light, the photons emitted from deeper in the crystal will be the result of the recombination of carriers, with the lifetime typical for the undamaged material. Importantly, this result supports the observation that simply monitoring the lifetime is not sufficient for determining the quality of a material, a conclusion that we drew also earlier, based on other data[31].

Figure 2b-MA-Bulk shows an increase of lifetime for medium LP (Figure 2b-MA-Bulk, top), where the PL increases along, along with a decrease of the lifetime for high LP (Figure 2b-MA-Bulk, bottom), where the PL intensity decreases. Note that the lifetime of the carriers emitted from unbleached parts of the material (background of image Figure 2b-MA-Bulk) changes between the surface (~ 6.5 ns) and the bulk (~ 5 ns). This effect can have two possible causes:

(1) solvent molecules (DMF and DMSO) that can interact strongly[32] with the Pb$^{2+}$ ions are incorporated into the HaP crystals as defects and behave as recombination centers, thus reducing the lifetime (mass spectroscopy gives evidence for solvent molecules in the single crystals; data not shown). Possibly solvent molecules at the surface were removed by evaporation but are still trapped in the bulk.

(2) there is a high enough density of shallow defects at the surface that can trap (minority) carriers, decrease the number of carriers that actually recombine per unit time, and increase the measured carrier lifetime (note that the trapped carriers eventually de-trap and recombine at longer times via band-to-band transitions). Such effects are known for other bulk semiconductors[33].

The Cs sample shows an effect similar to that found for the MA one for both bleached and background areas. Specifically we observe a reduction of PL lifetime of the background from ~10 ns at the surface to ~7 ns in the bulk. In this case, the solvent-related cause (option (1) above) seems even more probable; CsPbBr$_3$ crystals are grown in a DMSO solution and DMSO-Pb$^{2+}$ complexes are known to be formed in solution and to be crystallized in the presence of inorganic anions[34]. Concluding our investigation of the Cs sample at the surface for medium LP and in the bulk (Figure 2b-Cs), we find that bleaching reduces the lifetime. This is expected from the decrease in PL intensity, because a larger number of carriers recombine through faster, non-radiative, processes.

For completeness, we also report the observed values of lifetime for the more photo-luminescent, blue-shifted material obtained by bleaching the Cs sample at the surface with high LP. Figure 2b-Cs-Surface-High LP shows a reduction of the PL lifetime from 8-10 ns in the unbleached crystal to ~5.5 ns. Normally, a reduction of lifetime should correspond to a reduction of PL intensity, but that is not so in this case, probably because the blue-shifted material has a higher PL quantum efficiency than the perovskite background.

Finally, in Figure 2c we compare PL and FLIM images of the MA sample at different times after photo-bleaching in the bulk, showing similar lifetime and PL intensity changes during the healing process. As lifetime is proportional to the defect density in single defect models, simple PL can be used to follow the trap density over time. Given that we find that the PL self-healing decay is exponential, we can infer that a single (migrating) species is critical for the process. In the next two paragraphs we provide data that suggest that the species may be a (complex)bromine interstitial (high-LP) or methylamine-associated (medium-LP).

### 3. Results: Origin of self-healing in the bulk

In the preceding section we showed that in nearly all cases changes in PL intensity directly relate to PL lifetime changes. This suggests that both have a common cause, which could be a change of defect density. However, in section 2, we found that equivalent laser irradiation treatments affect the MA-, FA-, and Cs- perovskites differently. Moreover, we showed in section 2.1 that the A cation influences self-healing of the perovskites (cf. Figure 1). This indicates that the A cation is key in determining the stability of the perovskite. This conclusion points to a chemical basis for the differences, given the lack of a direct effect of the A cation on the electronic structure of the material.[35]

To identify the chemical origin of such differences in the bulk without resorting to destructive sample treatment, e.g. by investigating cross-sections (cf. ref. 11), we performed density functional theory (DFT) computations (for full computational details see SI-8). The following simulations focus on the deformations caused in the minimum energy structure of the pristine material by introducing a defect, as well as on the degree of charge localization around the defect.

The A cation is known to have a minimal direct effect on the HaP electronic structure.[35] However, it does influence the structure of the perovskite by changing the cell size and the symmetry, thereby indirectly affecting the defect chemistry of the material. Depending on the A cation, a similar defect can cause varying levels of distortion to the crystal, which can determine whether the defect will have long-term (meta)stability. The more distorted the structure, the less likely it is that the defect will persist.

A growing consensus points to halogen related defects, most notably interstitials, as possible traps and/or charge recombination centers in HaPs.[36–39] For that reason, we focus our efforts on interstitial bromide defects in all three systems. In the point defect model, an interstitial defect can have different charge states, depending on the energy of its charge transition level relative to the Fermi level of the material. Specifically, a Br interstitial defect can be either positively charged ($Br_i^+$), neutral ($Br_i^0$), or negatively charged ($Br_i^-$). Experimentally, HaPs are commonly found to be slightly p-type,[40,41] i.e., with a Fermi level that is closer to the level of the valence band maximum (VBM) than the conduction band minimum (CBM) and lower than the defect level. This suggests that a positively charged defect is the most probable one.

Using DFT, we calculated the minimum energy structure of a Br interstitial in each of its possible charge states, in all three systems. The $Br_i^+$ structures are presented in Figures 3a-c. For completeness, all other calculated minimum energy structures are reported in SI-8. In all three systems, the $Br_i^+$ defect produced a similar structure: the $Br_i^+$ relaxed to a position between two neighboring lattice Br anions, connecting two octahedral cages to create a *rigid, linear $Br_3^-$* structure with Br-Br distances of 0.255-0.259 nm. $Br_3^-$ anions are known constituents of solids, e.g. $CsBr_3$,[42] and other organic cation equivalents[43] or ionic liquids[44] (e.g., $MaBr_3$ and $FABr_3$)[15]. Linear poly-halide anion species generated by interstitial Br have also been observed experimentally in some crystal structures, e.g. KCl,[45,46] KBr,[45,46] RbBr,[47] and CsBr.[48]

We find that the room temperature (RT) orthorhombic $CsPbBr_3$ structure remains almost undistorted around the defect, whereas the RT cubic $MAPbBr_3$ and $FAPbBr_3$ structures are strongly perturbed. In particular, in $MAPbBr_3$ and $FAPbBr_3$, one of the bromine atoms has to shift strongly off the axis formed by the two Pb ions to which it is bound, in order to bind to the Br interstitial ion (see the displacement ($\Delta$) of the topmost Br atoms, marked in purple in Figures 3b and 3c (red arrows)). As a consequence, the interstitial Br atom and the second structural Br atom are pushed downwards and away from the latter's equilibrium position. While for $MAPbBr_3$ the $MA^+$ orientations remain relatively unchanged, in the case of $FA^+$ the presence of a Br interstitial defect causes a strong modification of the equilibrium positions of the nearby $FA^+$ ions (see ions indicated by light blue arrows, Figure 3c). In addition, reaching converged minimum energy geometries for the Br interstitial in $FAPbBr_3$ is much more difficult than in $MAPbBr_3$ and $CsPbBr_3$. This, coupled to the large structural distortions in the $FAPbBr_3$ system, indirectly indicates the improbability of a stable Br interstitial in the material. We can speculate that $FA^+$ ions, by way of their natural vibration, will "push" the interstitial Br ion out towards the next nearby equilibrium Br position (in the octahedral cage), with consequent displacement of the already-present Br ion that becomes an interstitial defect in the next unit cell. This process will repeat until the interstitial recombines with a vacancy or is otherwise annihilated. The DFT calculations therefore suggest *that the self-healing in the bulk of the $APbBr_3$ could be inversely related to the amount and localization of the deformation introduced by the Br interstitial.*[49] The stability of a Br interstitial defect is lowest for $FAPbBr_3$, whereas in $CsPbBr_3$ the defect can exhibit long term metastability.

It is important to note that it has recently become clear that the perovskite structure is not static, but rather is subject to large, anharmonic, polar fluctuations.[50–52] Moreover, these fluctuations have been shown to greatly affect the defect energy landscape.[53] Clearly, the static, relaxed structure can only represent an average picture. Because the static $FA^+$-containing structure is already greatly distorted even in the absence of dynamic effects, much more so than the Cs$^+$-containing one, it is reasonable to assume that these differences carry over to the dynamic, room temperature systems.

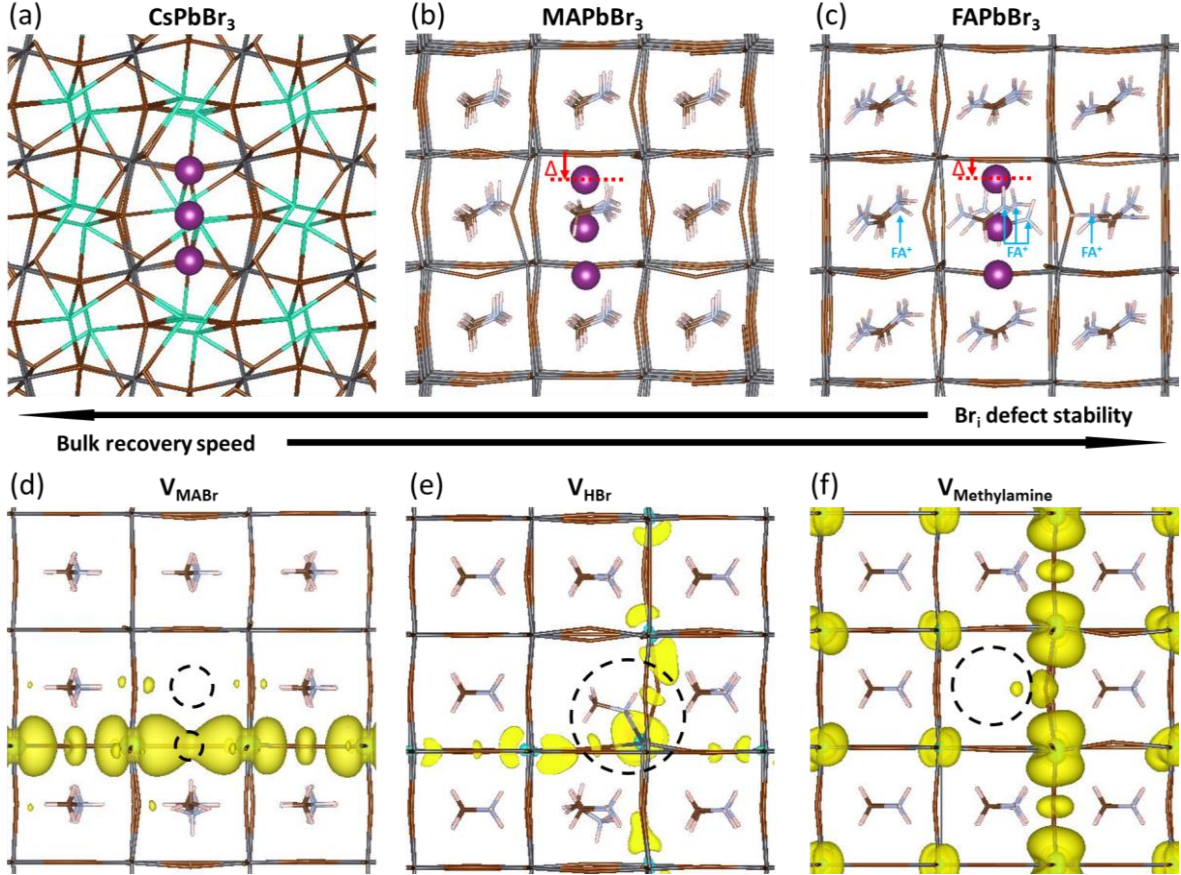

**Figure 3.** Minimum energy structures of (**a**) orthorhombic CsPbBr$_3$, (**b**) cubic MAPbBr$_3$, and (**c**) cubic FAPbBr$_3$, containing a Br$_i^\bullet$ defect, as well as minimum energy structures of MAPbBr$_3$ containing (**d**) V$_{MABr}$, (**e**) V$_{HBr}$, and (**f**) V$_{MeNH2}$. Br$_i^\bullet$ defects are marked in purple and vacancies are marked by dashed circles. Cs atoms are marked in green, Pb atoms in dark grey, and Br atoms in brown. The red arrow indicates the displacement of the Br atoms from the Br-Pb-Br axis. The light blue arrows indicate the FA$^+$ ions most affected by the defect. Yellow contours in (d)-(f) represent the partial charge density associated with the defect eigenvalue.

We remind that MAPbBr$_3$ exhibits increased PL brightness upon photo-bleaching in the bulk, whereas the two other bromide perovskites do not.

This necessitates further examination of the consequences of an MA-related defect. Free methylamine possibly forms in the bulk after dissociation of the methylammonium to methylamine and HBr. We therefore look into three neutral defect complexes involving methylamine and bromine: 1) an MABr vacancy, V$_{MABr}$; 2) an HBr vacancy, V$_{HBr}$; 3) a methylamine vacancy, V$_{MeNH2}$.

In Figure 3d, we show the structure and charge density associated with V$_{MABr}$. In this case, the charge density is localized along the Pb-V$_{Br}$-Pb axis. For the V$_{HBr}$ (Figure 3e), the charge density spreads two dimensionally, while for the V$_{MeNH2}$ (Figure 3f) the defect induces a smaller effect with an almost unchanged geometry and a rather delocalized charge density in all three dimensions. This indicates a negligible effect on the electronic structure and a very low likelihood of trap formation. We stress that for the V$_{HBr}$ we report a direct bond between the lone pair of the N atom of methylamine and the nearest Pb atom. When removing HBr (as well as MABr), two Pb-Br bonds are broken, creating a defect level in the gap. The additional electron density, due to the lone pair on the N atom in CH$_3$NH$_2$, partially screens the defect, making V$_{HBr}$ less harmful than V$_{MABr}$. However, V$_{HBr}$ and V$_{MeNH2}$ are not the thermodynamically favored structures. According to our calculations, the reaction: $V_{HBr} + V_{MeNH_2} \leftrightarrow V_{MABr}$ results in a 1.9 eV reduction in energy, indicating that the methylamine is likely to migrate back to recombine with the HBr "left behind", thereby restoring the MABr vacancy. *This ex-* plains why the more luminescent state, formed under photo-bleaching and comprised of $V_{HBr}$ and $V_{MeNH2}$, reverts to the original state, $V_{MABr}$, with time.

### 4. Results: Analysis of the Photo-damaged surface

Having investigated the origin of self-healing of halide perovskites in the bulk, we now turn to the consequences of the bleaching process on the surface. Unlike the bulk, surface effects are difficult to study computationally due to symmetry loss at the surface, a difficulty in capturing the interactions at an interface, and a multitude of possible surface terminations. However, many experimental methods are available for use in this case. In this study, we used AFM, SEM, EDS and XPS to characterize the damage at the (near)surface and to investigate changes of morphology and composition after the bleaching cycles.

Figure 2b-MA-Surface-High LP shows that MA samples exhibit a slightly more luminescent halo around the strongly bleached area. Because no bleaching was performed in that region, the modification of the PL has to originate from chemicals released upon bleaching of the surface. AFM imaging on a sample that had been bleached (High LP) for 100 consecutive times reveals the formation of small, possibly HaP particles (as the PL spectrum, not shown, remains that of MAPbBr$_3$). We suggest this is formed by the decomposition products, resulting from surface vaporization by the intense laser, that fall back on the sample after cooling. However, we cannot be certain of their composition because the PL spectrum can originate from the underlying crystal. Figure 4a shows AFM images, obtained 0.5 hr (i) and 1 day (ii) after the bleaching experiment of the area that showed a halo around the bleached circle. An increased number of bleaching cycles assists in locating the spot optically (Figure S6) and allows imaging of the halo area that formed away from the bleached area (details in SI.10). Comparing Figures 4a (i) and (ii), we infer that the material connecting the different particles, seen in (i), evaporates with time and is probably a viscous liquid, as suggested by its flattened shape. We attribute the presence of this liquid to the re-absorption of locally-condensed methylamine and NH$_3$, known to dissolve into the MAPbBr$_3$ to form a viscous phase.[54] The other products of decomposition (HBr, CH$_3$Br) do not interact with the solid and would not form such a phase. Because the vapor pressure of NH$_3$ at 25 °C (~10 bar)[55] is higher than that of CH$_3$NH$_2$ (~2 bar),[56] which is consistent with their boiling points (NH$_3$: -33.3 °C[57]; CH$_3$NH$_2$ : 6 °C[58]), we expect CH$_3$NH$_2$ to dominate re-condensation after vaporization. This argument can explain that we do not find any PL halo around the bleached areas for FA and Cs samples because their decomposition products cannot form such a liquid with the corresponding HaP.

Figure 4b-MA shows SEM images of the surface of the MA crystal. SE2 detector designates secondary electrons which are particularly sensitive to the surface morphology. Images with the In-Lens electron detector are more sensitive to the work function.[59,60] Clearly the crystal surface, where the material was bleached (with ~ 870 μW laser power), is strongly affected by the treatment.

The MA sample formed structures inside the bleached area. As shown above, some signs of the presumed liquid CH$_3$NH$_2$-rich phase are also visible by SEM outside the bleached zone (Figure 4b-MA). However, because SEM is performed in vacuum, the phase evaporates, with subsequent disappearance of the nanoparticles that were observed in the AFM images (Figure 4a). The InLens image shows an increased signal relatively to the background, indicating, as reported earlier,[15,59,60] a decrease in work function. This can be explained by the formation of electron-rich Pb$^0$ (work function of 4.0 eV,[61] compared to 4.7 to 5.1 eV for the HaP single crystals). The FA sample (Figure 4b-FA), even if less stable for a single, defect-inducing bleaching cycle, seems to be more resistant to multiple cycles of bleaching/laser ablation, as can be seen from the SE2-morphology image. The Cs sample appears to be as affected by the strong laser pulses as the MA one, with an evident change in morphology. Noticeably, for the Cs sample the In-Lens image shows a very strong increase of the signal in the bleached area, indicating a more substantial decrease of the work function than for the MA and FA samples. At the same time, the area around the bleach shows a decrease of signal, which corresponds to an increase of the work function. To explain these observations, we examined the surface through EDS and XPS scouting for variation in composition that would explain the above-mentioned observations. EDS analysis shows that photobleaching of the MA sample increases the Pb:Br ratio in the bleached area 1.4 times compared to the bulk. Considering that the EDS signal is collected down to a depth of ~0.5 μm, we can assume that the real Pb:Br ratio on the surface is actually much higher. No difference in composition, within the sensitivity of EDS, is found between the background and the halo-area around the bleached area. EDS analysis of FA samples, on the other hand, showed only a 1.15 times increase in Pb:Br ratio, which is consistent with less electron-rich Pb$^0$ formation. Cs samples show no change in the Cs:Pb:Br ratio but a Pb:O ratio of 1.4:1 is found indicating additional oxygen (see SI.12 – no O is found in the background by EDS).

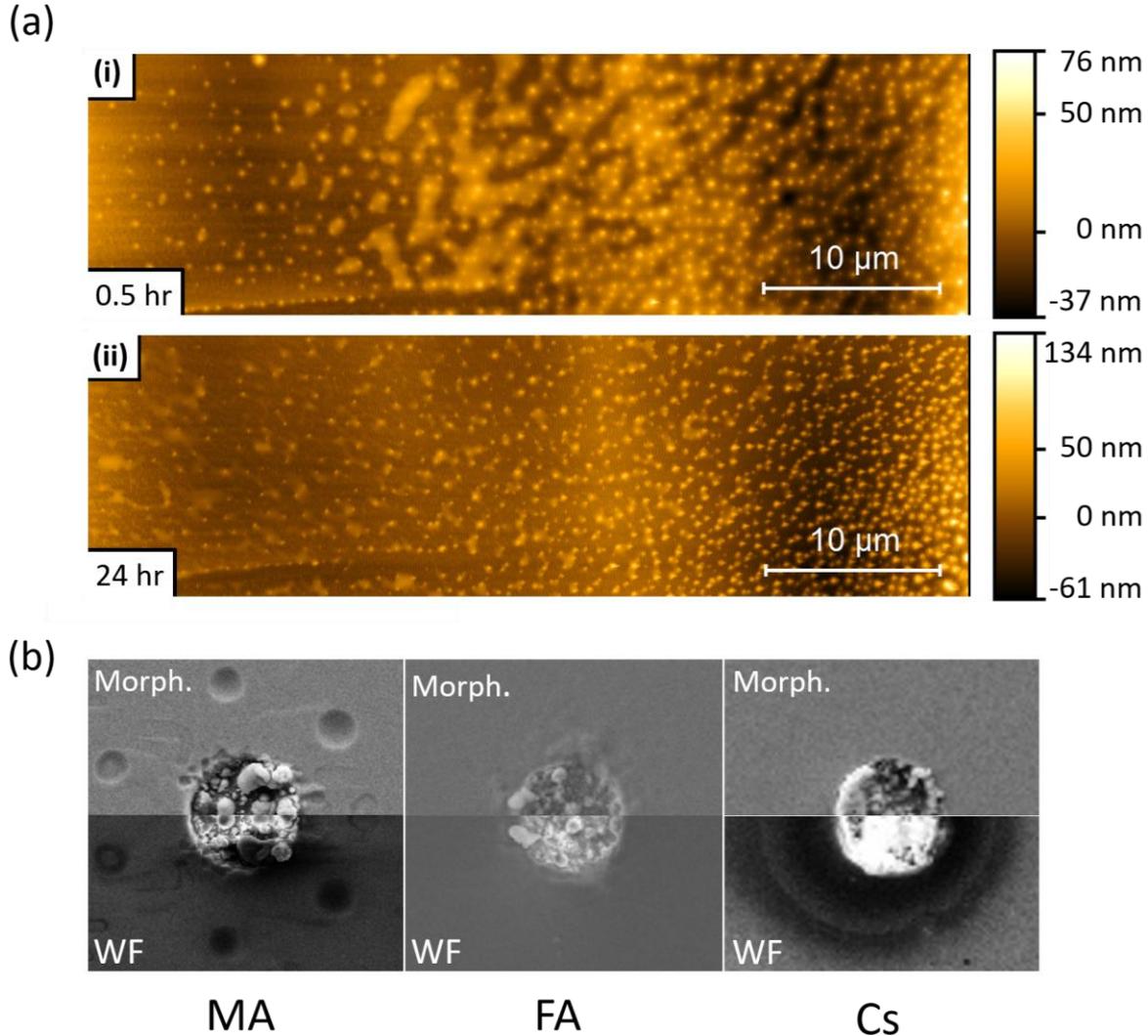

**Figure 4. (a)** AFM images of MAPbBr$_3$ single crystal surface 30 min **(i)** and 24 hours **(ii)** after 100 cycles of 1P bleaching with a laser power of 880 μW (LP). The AFM image is taken 20-30 μm from the bleached area (see Fig. S6), because the AFM tip could not make contact closer to the damaged area. The images show nanoparticles formed on the crystal surface and a semi-liquid phase between the particles that evaporates with time (compare top and bottom AFM images). **(b)** SEM images of MAPbBr$_3$, FAPbBr$_3$ and CsPbBr$_3$ single crystals after 100 cycles of 1P bleaching with LP. The images were taken with the SE2 detector to show the morphology and with the InLens detector to map variations in work function (whiter for smaller, darker for larger work function). The images are mirrored to facilitate comparison between them.

The analysis of a Cs sample, treated in the same way over a much larger (few mm$^2$) area, with XPS did not show the presence of any Pb oxide, as the Pb spectrum did not differ from that of a non-treated one. XPS revealed, instead, a ~ 15% increase of the Cs:Pb ratio on the surface. Since any sample is contaminated by oxygen and carbon and XPS cannot differentiate between traces of O coming from perovskite oxidation or from contamination, it is not possible to exploit it to determine O surface concentration. Nevertheless, as the depth analyzed by XPS is 1-2 orders of magnitude smaller than the EDS one, the two results are not contradictory. From the comparison of XPS and EDS, we conclude that Cs enrichment is limited to the surface region. The presence of oxygen is probably an indication of the formation of either lead or cesium oxides or, lead bromates. The formation of cesium oxides can explain the InLens-work function image of Figure 4b-Cs, because Cs oxides are known low work-function (~1 eV) materials.[39] We note that neither lead oxide nor lead bromate are stable in acid environment. MAPbBr$_3$ and FAPbBr$_3$ are reactive towards oxides due to the acidity of the A$^+$ cation and would transform to PbBr$_2$. PbBr$_2$ could then eventually transform to Pb$^0$ under strong illumination, a photoreaction known for PbI$_2$.[63] The possible formation of lead oxides or bromates would deplete the lead content of the original perovskite, leaving a Cs-rich volume that, as reported previously,[29] has a stronger and blue-shifted photoluminescence.

### 5. Discussion

The presented data show the differences in degradation and self-healing between surface and bulk and between MAPbBr$_3$, FAPbBr$_3$, and CsPbBr$_3$. These variations can be crucial in determining the stability of HaP materials and HaP-based devices. To understand the atomistic processes that lead to the reported results we first outline how the A cation influences self-healing on both the surface and in the bulk of HaPs. We then consider how the A cation influences the defect density in the HaPs. Finally, we stress how "proper" encapsulation has to take into account all processes we have identified, rather than being limited to the prevention of H$_2$O and O$_2$ penetration in the HaP-based device.

*Surface and Bulk Stability: The role of the A cation.*

In Figure 5 and in Table 1 we summarize all the information on APbBr$_3$ perovskites after photo-bleaching, provided by the current study. Closed circles represent degradation followed by self-healing processes. Red circles describe Br$_i$-related ones and green circles describe methylamine-related processes. Thicker lines correspond to faster processes. Arrows pointing vertically represent material exchange with the environment. Bulk and surface processes are separated by the dotted line, showing that self-healing dominates in the bulk and material loss dominates at the surface. Each of the indicated processes is supported by experimental data and computational theory as indicated by symbols (legend in the gray areas).

Analyzing the results for the FA HaP (left column in Figure 5), we see how FA heals completely and rapidly (thick red curved line) in the bulk, while showing mostly irreversible damage at the surface. DFT results showed how, due to steric effects, FA$^+$ ions around the Br$_i$• are displaced from their equilibrium position in the defect-free structure. It is reasonable to think that the found configuration is then energetically unfavorable. The system will try to revert to its defect-free form explaining the complete self-healing in the bulk. At the same time, surface results show that, once damaged, FA becomes more stable (more self-healing) than MA. Formamidinium does not decompose as methylammonium to an amine – acid couple, as does methylammonium, but decomposes to higher boiling point organic (aromatic) molecules i.e. sym-triazine[64,65]:

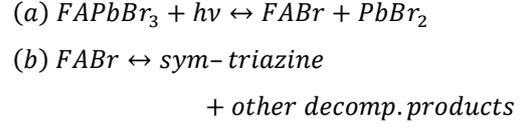

$(a)\ FAPbBr_3 + h\nu \leftrightarrow FABr + PbBr_2$

$(b)\ FABr \leftrightarrow sym\text{--}triazine + other\ decomp.products$

Examining the effects of the MA cation on the 2P bulk processes (center column in Figure 5), we observed two opposing transient phenomena (cf. Figure 1b): one increases the PL at medium LP (indicated by the green arrow), and another one decreases the PL, as in the FA and Cs samples, for which PL increases at higher LPs. The first effect is ascribed to passivation by methylamine, generated by decomposition of methyl ammonium. The second effect is ascribed to the chemistry of the Br$_3^-$ species the form of the interstitial Br defect (Br$_i$•), suggested by DFT.

Considering now the surface process, the MA perovskite decomposition proceeds as follows[12,13]:

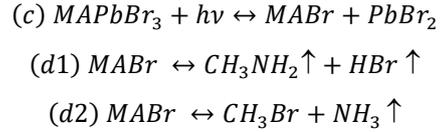

$(c)\ MAPbBr_3 + h\nu \leftrightarrow MABr + PbBr_2$

$(d1)\ MABr \leftrightarrow CH_3NH_2\uparrow + HBr\uparrow$

$(d2)\ MABr \leftrightarrow CH_3Br + NH_3\uparrow$

Above, we argued that the methylamine from reaction (d1) re-condenses on the crystal surface, eventually re-evaporating over long times. It has been shown[12] that CH$_3$NH$_2$ can be formed from MAPbBr$_3$ (films) under one sun illumination. After degradation, it is probable that, if the system is not in vacuum, at least some CH$_3$NH$_2$ is reabsorbed at the surface, yielding its known beneficial effect on the material[54,66]. This phenomenon can explain at least part of the reported improvement of the MAPbBr$_3$ properties after a short illumination period (light-soaking), reported in the literature[67,68].

In the bulk material, the products of reaction (d1) can be absorbed by the crystal without forming a gas: *methylamine*, as discussed above, can penetrate the crystal structure, *HBr* can dissociate into H$^+$, which can penetrate into the structure due to its small size, while Br$^-$ can be accepted as part of an interstitial defect. CH$_3$Br, however, is a hydrophobic molecule that does not readily react either with the perovskite or with NH$_3$. Its formation would build up extremely high pressure inside the crystal. This reasoning rationalizes that reaction (d2) is inhibited in the bulk by the law of mass action. However, in contact with an organic HTL, ETL or encapsulating polymer, $CH_3Br$ will be readily absorbed, pushing the reaction to the right and leading to irreversible decomposition, *even if the device is encapsulated*.

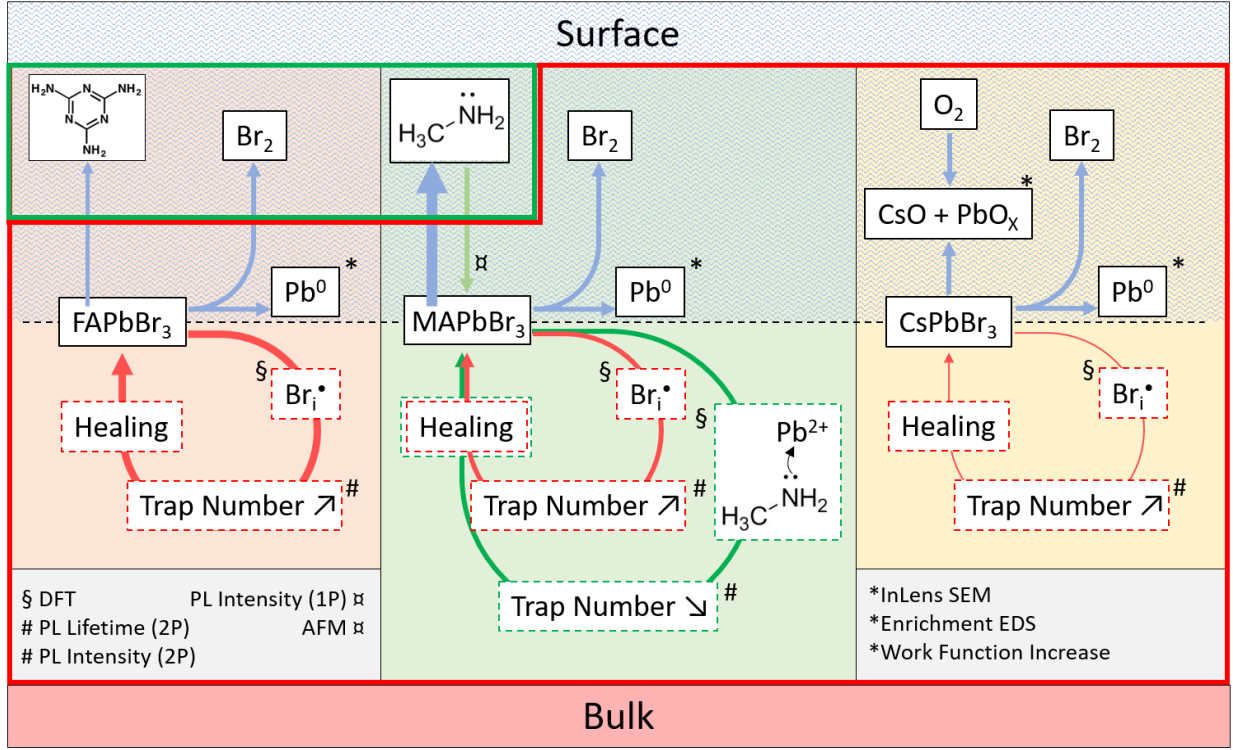

**Figure 5.** Scheme of the chemical processes described in literature (green rectangle) and in this work (red rectangle). Surface phenomena are reported in the upper row and bulk ones in the row below, under the dotted line. The thickness of the arrows (for the degradation on the surface and the self-healing in the bulk) corresponds qualitatively to the speed of the corresponding process. For further explanations, see text.

Comparing the chemistry of $MAPbBr_3$ and $FAPbBr_3$ we notice that reaction (c), involving $MAPbBr_3$, needs more energy to be pushed to the right by light (damage threshold 60 % of LP, full line Figure 1a) than reaction (a) involving $FAPbBr_3$ (threshold 45% LP, full line Figure 1c). Further MABr decomposition (d1-d2) needs less energy than that of FABr (b) (cf. Figure 1a, 1c, full lines). After the initial decomposition (b), the sym-triazine that formed, does not leave the bleaching site as easily as $CH_3NH_2$, HBr, $CH_3Br$ and $NH_3$, the MABr decomposition products (d1-d2), as it is has a higher boiling point and lower vapor pressure than those molecules. Therefore, sym-triazine pushes the equilibrium of the decomposition reaction to the left. The result is that reaction (b) is less likely to happen, more products of equation (a) remain and some self-healing happens at the surface to a greater extent than what is the case for its MA equivalent.

The situation is different for the $CsPbBr_3$ sample (right column in Figure 5). From the DFT computations we learn that in the bulk the formation of a $Br_3^-$ defect is not energetically expensive, as it does not significantly impact the crystal structure. This result explains, at least partially, why self-healing of the material in the bulk is slow and its stability is lower than that of its counterparts with organic A cation. At the surface, $CsPbBr_3$ has a greater ability to rearrange its structure as movements of atoms on the surface are less sterically hindered. Because of this, the migration of the products of decomposition (excluding the oxides) is easier than in the bulk once the decomposition happens. Notably, at the surface (Figure 1e) the minimum light intensity at which the Cs sample is damaged is lower than for the MA one. Thus, intrinsically, CsPbBr$_3$ is less stable than the MA analog. The LP threshold of damage (cf. Figure 1a, 1c, 1e full lines) can be viewed as a proxy for the activation energy of the decomposition reaction. We can then conclude that the activation energy for Cs decomposition on the surface is less than that for MA decomposition and similar to that for FA decomposition. Still, its stability in normal use can be attributed to its better (less problematic) self-healing properties at the surface (cf. dashed lines in Figure 1e with dashed lines in Figure 1c and 1a).

| Cation | Location | Damage | Healing | Phenomenon | Support |
|---|---|---|---|---|---|
| MA | Surface | Threshold | Minimal | Halo – MA condensation<br>Pb / Br increase | AFM – SEM<br>EDS |
|  | Bulk | Progressive | Fast | Increase in PL | DFT |
| FA | Surface | Threshold | Partial | Pb / Br increase | EDS |
|  | Bulk | Progressive | Very Fast | - | DFT |
| Cs | Surface | Threshold | Semi-complete | O-containing product(s)<br>Cs / Pb increase | EDS<br>XPS |
|  | Bulk | Progressive | Slow | - | DFT |

**Table 1**. Summary of the damage and self-healing and additional phenomena reported in this work for the $APbBr_3$ perovskites.

Interestingly, only for CsPbBr$_3$ do we find formation of a non-healable form with more intense PL inside the bleached area in the 1P-surface experiment at high LP (Figure 1e, Figure 2a and Figure 4b). This result is explained by the reaction of CsPbBr$_3$ with oxygen, as shown by EDS, which indicates formation of an O-containing product. XPS shows formation of a Cs-rich surface. This decomposition pathway creates a PL increase, only if the surface of the material is exposed to O$_2$ (which is the case under our experimental conditions) *and* to light. Still, this strong bleaching-induced modification of the surface does not improve the PL lifetime, despite increasing its intensity (Figure 2b-Cs-High LP).

Over-all, surface decomposition of CsPbBr$_3$, even if starting at lower energies than in the MA case, is less impactful than for FA and MA, probably because only a small amount of volatiles is actually created and most of the material can actually re-form (and, likely, quicker than in the bulk, as steric constraints are relaxed).

By integrating the data on the three HaPs that we studied and their analyses, we can consider the case of mixed A cation perovskites MA$_x$FA$_y$Cs$_{1-x-y}$PbBr$_3$: Cs helps to stabilize the surface of the material, as CsPbBr$_3$ cannot decompose further. Under optimal working conditions, the Cs, MA and FA HaPs should, after exposure to sunlight, starts achieve a state where most of the surface is Cs-rich, some methylamine is released and able to migrate into the bulk to heal defects and the formamidinium grants the long-term stability thanks to its superior bulk self-healing properties. We can speculate that at steady state (of illumination), the A cation distribution does not remain uniform and evolves towards a Cs-enriched stabilized surface and an FA-enriched self-healing-promoting bulk. In the future, this model can be tested by, e.g., ToF-SIMS experiments of films or crystals, exposed to light for long times and such experiments are planned.

Summarizing, each A cation influences the chemistry and self-healing of the perovskite differently. When the A cations are mixed, each of them can synergistically contribute to the overall stability and performance of the material. This synergism is in addition to configurational entropy gains of the system, due lack of order in A cation distribution.

*The importance of self-healing for bulk defect density*

The FA$^+$ cation plays here the role of keeping the defect density low. As the defect density is crucial for HaP-based solar cells, we can understand why FA$^+$ is the most abundant cation in MA$_x$FA$_y$Cs$_{1-x-y}$PbBr$_3$ based solar cells.

At equilibrium a material's self-healing kinetics are connected to its defect density.[69] For every defect, we consider that the defect density is determined by the equilibrium between formation and destruction (by healing) of the defect:

$$(e) \quad APbBr_3 \underset{k_{self-healing}}{\overset{k_{formation}}{\rightleftarrows}} defect$$

For any defect, the equilibrium defect density [*defect*] is:

$$(f) \quad [defect] = K_{eq} \cdot [HaP] = \frac{k_{formation}}{k_{self-healing}} \cdot [HaP]$$

where [HaP] is the concentration of the perovskite, which is effectively constant for any relevant [*defect*]. From this we deduce that the [*defect*] is inversely proportional to the self-healing kinetic constant. We can then expect FAPbBr$_3$ to have the lowest defect density, as its self-healing kinetics are fastest. The effect is particularly relevant during solar illumination. In this case, loss of efficiency is commonly attributed to the formation of defects that are continuously created by illumination.[16,70] The self-healing process continuously repairs (making them disappear) the defects, dictating a steady-state defect density that normally will be higher than the equilibrium one. The faster the recovery kinetics, the closer the steady-state defect density is to the equilibrium one as dictated by thermodynamics. Therefore, self-healing protects the material from damage, enhancing its chemical stability and decreasing the steady state density of performance-damaging defects.

*The relation between self-healing and encapsulation*

For the self-healing to take place, any material loss has to be avoided to minimize the occurrence of reactions like *(a)* and *(c)*. For this reason, it is now necessary to redefine what "proper" encapsulation is in the case of HaPs. There are many reports on how classical encapsulation significantly improves the stability of halide perovskite solar cells. This is not surprising, as the most obvious effect of encapsulation is protecting the material(s) against chemical attacks (i.e. $H_2O$, $O_2$) from the ambient. Yet, encapsulation should serve another critical purpose for stability: it should impede any loss of decomposition products, resulting from damage of the perovskite. Containing these products will, by the mass-action law, convey stability. An effect explaining very recent results[71]. The self-healing that we demonstrated to occur in bulk perovskites requires any decomposition product to be readily available, to enable reversal of the decomposition reaction. This condition is clearly not fulfilled at a surface that is open to the ambient, with the consequence that self-healing will be limited. In most reported studies, the "escape" ways for decomposition products in HaP-based solar cells are not controlled. For example, an iodine (or bromine) molecule forming at the surface of a solar cell may well find its way between grain boundaries or through the larger than molecular spaces between the perovskite layer and the HTL (or in the porous, if not completely filled ETL as in case of mesoporous $TiO_2$). Eventually the molecule can get away from the HaP surface and, for example, accumulate in open spaces. In such case, even if the product of the decomposition has not reacted with the ETL or HTL, but ends up spatially far from the decomposition site, it will not be available for self-healing.

This problem is particularly important for HTL materials. Most HTLs used today, are not resistant to acids (HBr, HI), bases (methylamine, formamidine), organic molecules ($CH_3I$, $CH_3Br$), the types of organic solvents, used during HaP preparation, and/or halogens ($Br_2$, $I_2$), each of which may be formed/released by halide perovskite degradation. Most organic HTL materials, such as spiro-OMeTAD[72], poly-thiophenes[73] (P3HT, PTB7, etc.) and poly-triarylamine PTAA, will eventually absorb or react with some or all decomposition products. With a few exceptions, such as PCBM[74] (phenyl-C61-butyric acid methyl ester), ETLs are inorganic oxides, which will mostly be more stable than organic ones.

To conclude our results also point to one reason why hybrid 3D-2D perovskites can have enhanced stability and optoelectronic properties. The organic layers of these perovskites can prevent escape of any 3D decomposition products. Such effect would be in addition to passivation of defects on the surfaces of the 3D perovskites, by adding the 2D layers.

**Conclusions**

We compared and contrasted experimentally bulk and surface damage of $APbBr_3$ halide perovskites and analyzed the results, using thermodynamic, kinetic, chemical, and physical considerations. We demonstrated how the intrinsic stability of the perovskites, linked to thermodynamics, is significantly greater than their kinetic stability, thanks to self-healing.

For applications, analyses of the degradation processes, whether thermal or light-induced, at the surface and in the bulk are very relevant, as they point to specific roles for each of the three (major) cations that are found empirically to be required for devices with optimal PV performance. FA provides superior healing properties in the bulk of the material, reducing the bulk defect density. When photoexcited in the bulk, MA decomposes to methylamine and HBr (equations (c) and (d1)). Methylamine can migrate and passivate defects while $H^+$ remains at the site of the $MA^+$ ion without serious effect on the local electronic properties. This causes an increase of PL (and probably in photo-conversion) efficiency, increasing solar cell performance. Finally, $Cs^+$ helps to stabilize the surface of the material.

In terms of fundamentals our results lead us to hypothesize about the question of *why HaPs have such good optoelectronic, and particularly PV properties,* expressed by (a) relatively long charge carrier lifetimes and diffusions lengths. These qualities can be traced to (b) low densities of electronically and optically active defects, i.e., those that affect measurable properties. As (a) will be qualitatively inversely proportional to (b), if the defects are subject to self-healing, their density will be small. Taken to the extreme, self-healing becomes the fundamental cause for the excellent opto-electronic properties of the HaPs, because all derive from the self-healing. Even if self-healing is not the sole cause, it likely is a major reason for the HaPs' excellent properties.

We stress how from our results we can infer that fulfilling the conditions for self-healing is necessary for long-lasting stability of devices made with HaP materials. In particular, apart from perfect encapsulation, the ETL and HTL have to be chemically unreactive towards the decomposition products of the perovskites in order to create "bulk-like" conditions for the HaP part of the interfaces. Likely in the most stable devices that were reported, such effect was an unintended consequence, analysis of which can help deciding on device architectures for commercialization.

**Acknowledgements.**

We thank Dan Oron (Weizmann Inst. of Science, WIS) for many fruitful and stimulating discussions and Ilario Gelmetti for fruitful discussions and clarity check of the text. Y.R. acknowledges partial support from a Perlman research grant for student-initiated research in chemistry. A.V.C, L.K, G.H. and D.C. acknowledge support by the Minerva Centre on self-repairing systems for energy & sustainability. D.C. also thanks the Yotam project of WIS' Sustainability And Energy Research Initiative, SAERI, for partial support. L.K. holds the Aryeh and Mintzi Katzman Professorial Chair. D.R.C. thanks the WIS'SAERI for a PD fellowship.

# SUPPLEMENTARY INFORMATION

# The pursuit of stability *in halide perovskites*: the monovalent cation and *the key for surface and bulk self-repair*

D R Ceratti[1]*, A V Cohen[1], R Tenne[2], Y Rakita[1], L Snarski[1], L Cremonesi[3], I Goldian[4], I Kaplan-Ashiri[4], T Bendikov[4], V Kalchenko[5], M Elbaum[6], M A C Potenza[2], L Kronik[1*], G Hodes[1*], D Cahen[1]*

[1]Weizmann Institute of Science, Department of Materials and Interfaces, 7610001, Rehovot, Israel.
[2]Weizmann Institute of Science, Department of Physics of Complex Systems, 7610001, Rehovot, Israel.
[3]Department of Physics and CIMAINA, University of Milan, via Celoria, 16 20133, Milan, Italy.
[4]Weizmann Institute of Science, Department of Chemical Research Support, 7610001, Rehovot, Israel.
[5]Weizmann Institute of Science, Department of Veterinary Resources, 7610001, Rehovot, Israel.
[6]Weizmann Institute of Science, Department of Chemical and Biological Physics, 7610001, Rehovot, Israel.

*davide-raffaele.ceratti@weizmann.ac.il , david.cahen@weizmann.ac.il,
gary.hodes@weizmann.ac.il, leeor.kronik@weizmann.ac.il

# Experimental section

## SI.1 Single Crystal Synthesis

MAPbBr$_3$, FAPbBr$_3$ and CsPbBr$_3$ single crystals were grown by the method reported in the supplementary information of ref. [1]. We add here details of the synthesis that help to obtain larger crystals with the anti-solvent method than before. To that end we cover the container with the perovskite precursors with filter paper. This avoids unwanted dropping of anti-solvent that condensed on the crystallization chamber walls. Any anti-solvent drop falling directly in the precursor solution induces the formation of multiple seeds, which reduces average crystal size. The verification of the crystallinity of the single crystals was done as reported in the supplementary information of ref.[1].

## SI.2. Two-photon (2P) vs one-photon (1P) microscopy

In 2P microscopy we can assess the bulk instead of the surface of a sample. Figure S1a illustrates how supra-bandgap laser light (488 nm) is absorbed in the first layers of the halide perovskite single crystal due to the high value of the 1P absorption coefficient of the perovskite. 2P excitation is obtained using sub-bandgap light. To get sufficient intensity to yield a measurable, useful density of photons with twice the energy of the exciting ones at the focal point inside the crystal, we need to use a pulsed laser. We use excitation from a pulsed laser at 800 nm, 1.55 eV, an energy well below the bandgap of the bromide halide perovskites, i.e., the single crystals are transparent for these photons, which, thus, can penetrate the crystal. Because of the difference in refractive index between the air and the crystal, the focus of the microscope objective is distorted with consequent positive aberration. Despite this, the 800 nm light remains focused enough for 2P absorption, which happens where the laser intensity is most concentrated. We reported in ref. [1] the shape of the focus and intensity distribution of the laser in the 2P microscope both at the surface and in the bulk. In Figure S1 we show the shape of the 1P focus and laser light intensity distribution as used for this experiment. While the same objective serves the 2P and 1P experiments, the Gaussian profile of the 1P laser is narrower than the objective back entrance. The effective numerical aperture of the objective was determined to be NA=0.505 by mapping the power profile of the laser at different distances from the objective. This number was used as input to obtain the profile reported in Figures S1c and S1.d.

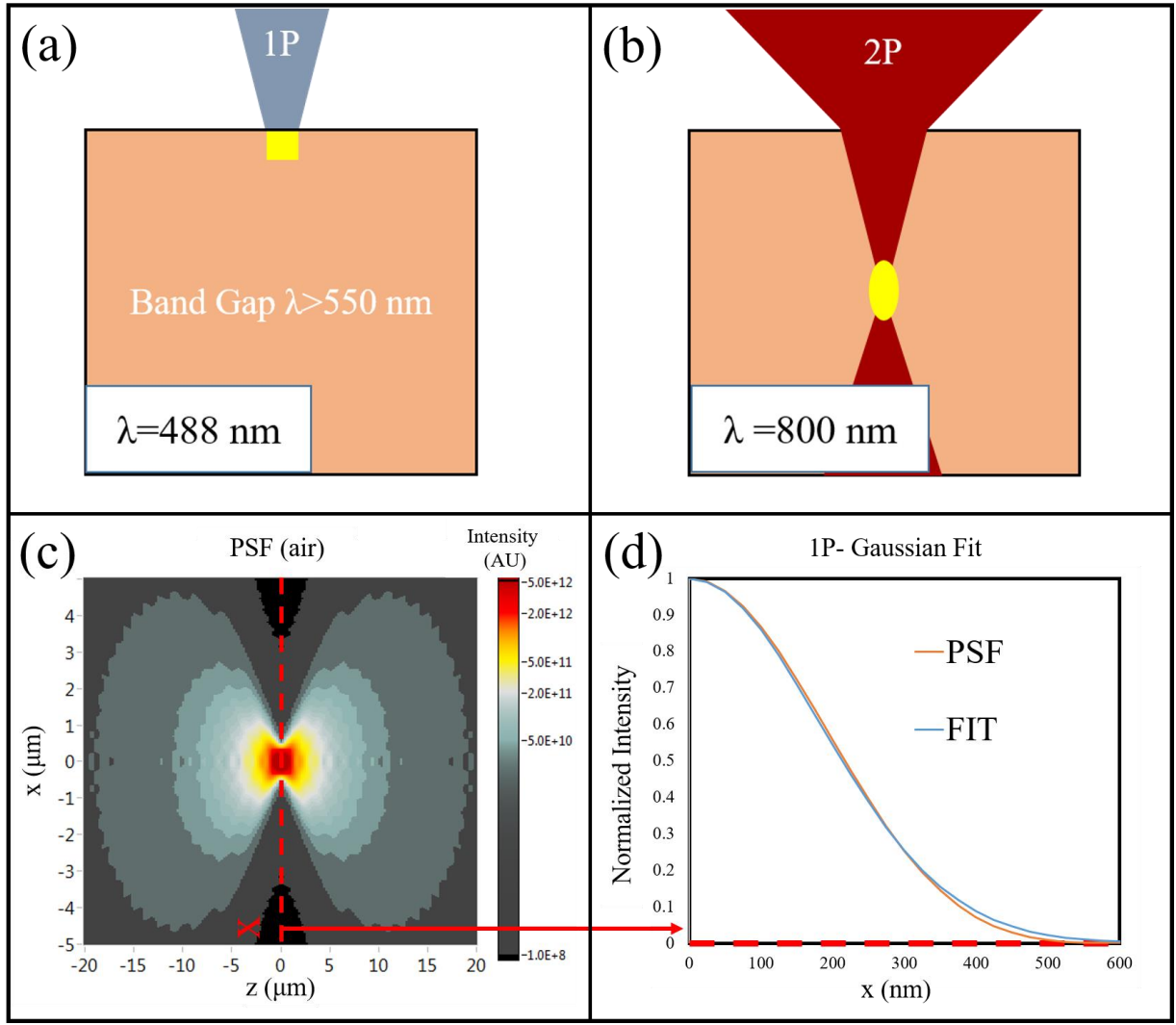

**Figure S1 :** Schematics of (a) normal (one-photon, 1P-488 nm–supra-bandgap) and (b) two-photon, 2P (800 nm-sub-bandgap) excitation (for APbBr$_3$ materials) microscopy and photobleaching. While the supra-bandgap light (1P) is absorbed by the halide perovskite within a few 100 nm, exponentially decreasing from the surface, (left), sub-bandgap light (2P) penetrates the crystal and is absorbed within the crystal. This behavior depends on the non-linear dependence of 2P absorption on light intensity. Due to the material's high refractive index, the focus of the light inside the crystal is shifted to deeper inside the material. This changes the energy deposition by 2P absorption. Accurate calculations are needed to calculate this aberration effect, and are reported in the following section. **(c)** I-P intensity distribution around the focus (point spread function - PSF) of an objective with NA=0.75 and effective NA=0.505. The 488 nm laser profile is indeed smaller than the objective back aperture diameter. (d) Gaussian fit of the PSF on the x axis. The PSF is normalized to its maximum value. The fitted σ of the Gaussian is 181.3 nm.

## SI.3. FRAP procedure

A FRAP experiment is usually composed of 3 steps: 1- obtain the PL image before damage (Figure S2-A), which gives the reference PL signal. 2- cause damage (bleaching) with high laser power (Figure S2-B); 3- monitor PL recovery over time (at imaging intensity, which is << than bleaching intensity) (Figure S2-C-D). The intensity of the bleaching can be varied. The maximum laser power that is used defines the maximum bleaching power. Commonly the laser

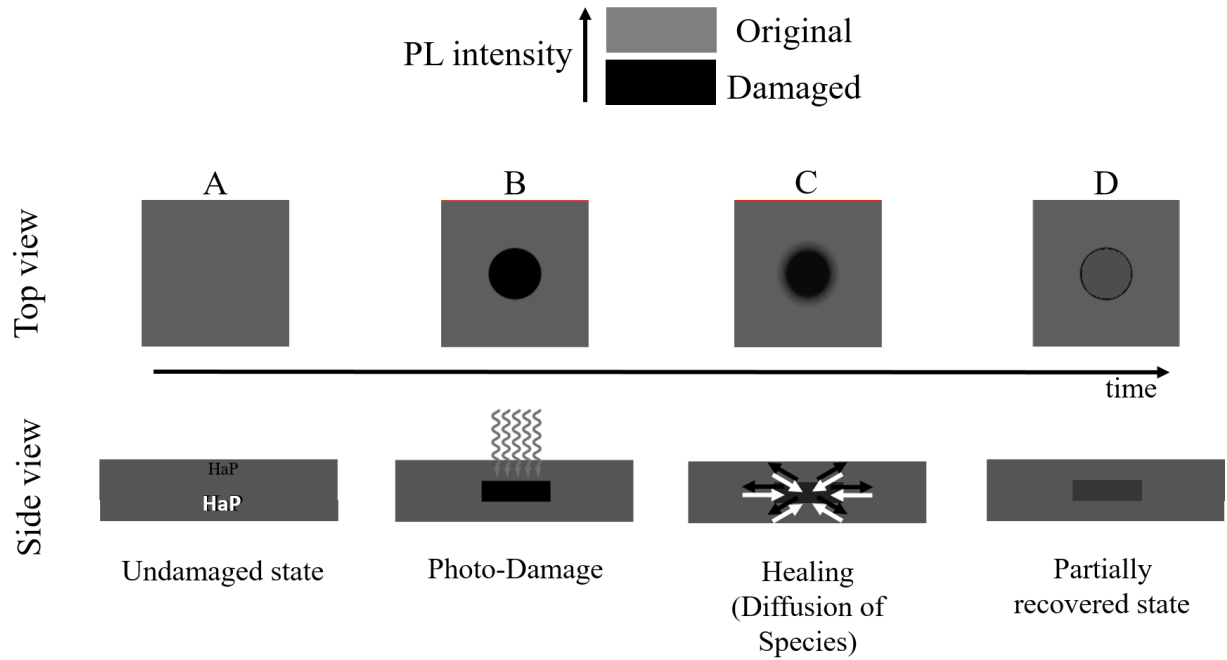

**Figure S2** : The successive steps of the FRAP experiment; cf. SI.3 text for explanation.

power can be tuned with an accuracy of 0.1% of the maximum laser power.

## SI.4. Definition of the value of $\varepsilon_{sun}$.

$\varepsilon_{sun}$ is defined as the energy per unit of volume (J/m$^3$) that is absorbed at and near the surface by MAPbBr$_3$ upon exposure to AM1.5G solar radiation in one second. In the following we use MAPbBr$_3$ as example and note that results for FAPbBr$_3$ and CsPbBr$_3$ will differ slightly values their absorption is not exactly the same as that of MAPbBr$_3$.

$$(a)\ \varepsilon_{sun} = \left(\frac{dI_{sun}}{dz}\right)_{surf} \cdot \Delta t_0$$

$$(b)\ \left(\frac{dI_{sun}}{dz}\right)_{surf} = \int_{200\ nm}^{1600\ nm} \alpha_{MAPbBr3}(\lambda) * I_{Sun}(\lambda)\ d\lambda$$

with $\alpha_{MAPbBr3}(\lambda)$ the wavelength-dependent 1P absorption coefficient of MAPbBr$_3$[2] and $I_{Sun}(\lambda)$ the wavelength-dependent solar intensity (from[26]), where 200 and 1600 nm are taken to (over)assure that all solar radiation is included. $\left(\frac{dI_{sun}}{dz}\right)_{surf}$ is the solar power/unit volume, absorbed at the surface in the very first layer of material; with $\Delta t_0$ = 1 sec, $\varepsilon_{sun}$ = 2.2·10$^9$ J • m$^{-3}$. This value defines the density of solar energy deposited in a MAPbBr$_3$ crystal or thin film in the top layers in one second and can be considered as a volume-normalized dose.

Note that AM 1.5 radiation, integrated between 200 (again to (over)assure that all radiation is included, as most UV radiation < 300 nm is absorbed by ozone) and 530 nm (corresponding approximately to the onset of MAPbBr$_3$ absorption) is $\int_{200\,nm}^{530\,nm} I(\lambda)\,d\lambda = 300\,W \cdot m^{-2}$. Because of the high 1P absorption coefficient of MAPbBr$_3$ at 500 nm, $\sim 10^7\,m^{-1}$, the density of energy actually absorbed by the material quickly decays while penetrating into it. One can make an approximate evaluation considering that all the 300 W • m$^{-2}$ of photon energy gets absorbed in the first 200 nm of the crystal (an overestimate as ~ 90% of the absorbable light is absorbed in that thick a film). By this simple calculation we obtain $\varepsilon_{sun}$ = 1.5·10$^9$ J • m$^{-3}$. However, for our discussion we keep the formally correct value of $\varepsilon_{sun}$ = 2.2·10$^9$ J • m$^{-3}$.

## SI.5. Laser intensity and deployed energy in 2-P and 1-P experiments.

In the Supplementary Information of ref.[1] we discuss the amount of aberration of 800 nm light inside a MAPbBr$_3$ single crystal. The evaluation of the aberration and, consequently, of the real absorption of light and energy by the bleaching treatment, depends on various instrumental parameters. Calculations that take into account all the critical parameters for 2P absorption were performed and the results reported in ref. [1].

Following the same notation as that used in the SI of [1] we define the value for $Sup_{488nm}$. As before[1] the pixel size is smaller than the light resolution. Because of this when the laser is directed to the following pixel, the preceding one is still in the focus of the objective and is exposed to laser light. The $Sup_{488nm}$ factor is a multiplicative factor that corresponds to the ratio between the actual absorbed energy in the area of the pixel and the energy that would be absorbed in the area of the pixel if the laser would have been on when pointing at the pixel only and then switched off when pointed at the following pixel. In our experiments the pixel size is 250 nm, $\Delta t$ = 8.0 μsec and $\sigma$ the standard deviation of the Gaussian profile (equal to 181.3 nm) as imposed by the instrument. Considering these parameters we obtain a value of $Sup_{488nm} = 3.30$. Using the values obtained by the PSF (point spread function cf. SI.2) simulations and considering all the mentioned conditions and the maximum incoming time-averaged laser power equal to 880 μW for the 1P microscope and 120 mW for the 2P microscope we have:

For the 1P case

(a) $W_{sun}^{1P} = \dfrac{\frac{Max(PSF)_{1P}}{2}*LP_{488}*\Delta t_{1P}*\alpha_{MAPbBr3}^{488}*Sup_{488nm}}{\varepsilon_{sun}} = 670\ (\varepsilon_{sun})$

where $W_{sun}^{1P}$ is the energy density deposited by one bleaching cycle at 100% of the laser power ($LP_{488}$); $Max(PSF)_{1P}$ is the maximum value of the Pointing vector as calculated by the simulation in SI.2 (the division by 2 comes from the averaging over the time of its oscillating value); $\Delta t_{1P}$ is the pixel dwell time; $\alpha_{MAPbBr3}^{488}$ is the 1P absorption coefficient of MAPbBr$_3$ at 488 nm; $Sup_{488nm}$ was defined above and $\varepsilon_{sun}$ is the above-mentioned value of the density of energy absorbed at the surface in one second by MAPbBr$_3$. At lower percentage of the laser power (X%) a value equal to $\dfrac{X}{100} * W_{sun}^{1P}$ is deposited on the sample.

For the 2P case, instead, we have

(b) $W_{sun}^{2P} = \dfrac{\frac{3}{8}Max(PSF)^2{}_{2P}*LP_{800}^2*\frac{\tau_{rep}}{\tau_{pulse}}*\Delta t_{2P}*\beta_{MAPbBr3}^{800}*Sup_{800nm}}{\varepsilon_{sun}} = 770\ (\varepsilon_{sun})$

with $W_{sun}^{2P}$, the energy density deposited by one bleaching cycle at 100% of the laser power ($LP_{800}$); $Max(PSF)_{2P}$ the maximum value of the Pointing vector as calculated by the simulation in[1] (the multiplication by 3/8 comes from the averaging over time of the square of its oscillating value) ; $\Delta t_{2P}$ is the pixel dwell time; $\beta_{MAPbBr3}^{800}$ is the 2P absorption coefficient of MAPbBr$_3$ at 800 nm; $Sup_{800nm}$ is as $Sup_{488nm}$, the multiplication factor that was defined above; $\tau_{rep}$ is the inverse of the pulse repetition rate of the laser (12.5 ns); $\tau_{pulse}$ the duration of the pulse (140 fs) and $\varepsilon_{sun}$ the above-mentioned value of the density of energy, absorbed at the surface in one second by MAPbBr$_3$. At a lower fraction of the laser power (X%) a value equal to $\left(\dfrac{X}{100}\right)^2 * W_{sun}^{2P}$ is deposited in the sample to take the quadratic dependence of the 2P absorption on the laser power into account.

In the 1P case, actually, the absorbed energy is captured by the (hot) carriers which loose part of their energy (0.2 eV) by scattering heating the system. The carriers then recombine ether through traps or through bimolecular recombination producing more heat (2.34 eV). Therefore, the volume in which the heat is released has the potential to be larger than the volume of absorption of the light accordingly to the carriers' diffusion length. Halide perovskite single crystals have shown to have diffusion length of the order of multiple μms with values reaching[3] the 175 μm and 3 mm. Taking the most unfavorable case found in literature[4] (which gives rise to the highest temperature), we consider the diffusion length of the carriers to be 3 μm. We can then consider the energy absorbed from the material to be released with a distribution roughly Gaussian with inflection point at 3 µm from the maximum absorption in the x,y,z direction. In the 1P case we can consider the 3 µm form the surface and from the center of the laser spot.

Concretely, if light is focused on a 272 nm × 272 nm (diffraction limit) spot and completely absorbed in around 200 nm depth, heat is released to x,y distance and depth that are actually 15 times larger. This means that the previously calculated $W_{sun}^{1P}$ should actually be divided for a factor 15[3]. It is our opinion that the actual diffusion length in our crystal is actually smaller than 3 µm. Measuring its exact value in the relevant conditions (partially illuminated, with non-uniform carrier concentration, sample) is not trivial and is out of the scope of this article.

In the 2P case a similar effect of diffusion of the energy should be considered. However, the 2P absorption is more diffuse in the volume. This means that the dilution of the energy density calculated in the 2P case is less pronounced than the one calculated in the 1P case.

This effect provides a likely explanation for the discrepancy between the $\varepsilon_{sun}$ values at which we see the first damage in the 2P and 1P cases bringing the actual value of energy density in the 1P case down to the 2P case (if not lower).

## SI.6. Evaluation of the temperature during the FRAP experiment.

The temperature rise upon bleaching can be calculated knowing the thermal parameters of the halide perovskites. The light deploys energy with a distribution that is dependent on light focusing and light concentration. In the previous section, we calculated the energy released by the laser during a bleaching cycle. We also mentioned that considering the distribution of energy release equivalent to the light absorption distribution is an inexact approximation. In fact, the charges recombine with a Gaussian profile with inflection point of at least 3 µm.

From the diffusion equation

$$(c) \quad \Delta^2 T + \frac{HS}{\rho\, C_p} = \frac{1}{\alpha} \frac{\partial T}{\partial t}$$

With T as temperature and α the thermal diffusivity (with k the thermal conductivity, ρ the density and $C_p$ the heat capacity of the material)

$$(d) \quad \alpha = \frac{k}{\rho\, C_p}$$

Considering the values for MAPbBr$_3$ of heat capacity[5] (180 J•K$^{-1}$•mol$^{-1}$ which divided for the molecular weight (MW=479 g•mol) becomes 0.375 J•K$^{-1}$•g$^{-1}$), the mass density[6] (ρ=3.83 g•cm$^{-3}$), heat conductivity (measured[7] to be 0.44 W•m$^{-1}$•K$^{-1}$ comparable with the one of water) we have a product $\rho\, C_p = 1.44 * 10^6\ J \cdot K^{-1} \cdot m^{-3}$. So $\alpha = 3.05 * 10^{-5}\ m^2 \cdot s^{-1}$.

If we consider all the heat injected into the system instantaneously at time t' from a single point of coordinates x', y' and z' we have as a solution of the heat equation:

$$(e) \quad \Delta T(x,y,z,t) = \frac{Q_{instant}}{\rho\, C_p} * \frac{1}{(4\pi\alpha(t-t'))^{\frac{3}{2}}} Exp\left(-\frac{((x-x')^2+(y-y')^2+(z-z')^2)}{4\alpha(t-t')}\right)$$

If the heat is not released instantaneously but over time then it is possible to rewrite the equation as (with the time t'=0 the for the first released heat):

$$(f) \quad \Delta T(x,y,z,t) = \int_0^t dt' \frac{Q(t')}{\rho\, C_p} * \frac{1}{(4\pi\alpha(t-t'))^{\frac{3}{2}}} Exp\left(-\frac{((x-x')^2+(y-y')^2+(z-z')^2)}{4\alpha(t-t')}\right)$$

The heat can also be released not in one point only but can have a distribution:

$$(g) \quad dT(x,y,z,t) = \int dz' \int dy' \int dx' \int_0^t dt' \frac{Q(x',y',z',t')}{\rho\, C_p} *$$
$$\frac{1}{(4\pi\alpha(t-t'))^{\frac{3}{2}}} Exp\left(-\frac{((x-x')^2+(y-y')^2+(z-z')^2)}{4\alpha(t-t')}\right)$$

We can describe the heat source distribution as Gaussian with $\sigma$ the flex point distance and P the total energy deployed per unit of time.

$$(h) \quad Q(x',y',z') = \frac{P}{(2\pi\sigma^2)^{\frac{3}{2}}} Exp\left(-\frac{(x'^2+y'^2+z'^2)}{2\sigma^2}\right)$$

This distribution of heat can be considered as the heat originated from a point that already had some time $\tau$ to diffuse (this is because Gaussians evolve into Gaussians and a point injection of heat can be seen as heat injected with Gaussian distribution into an extremely small, point like, volume).

$$(i) \quad \tau = \frac{\sigma^2}{2\alpha}$$

Considering $\sigma = 2.5\ \mu m$ we would have a time $\tau = 10^{-7} s$. The heat distribution at the end of the bleach $t > t_{bleaching}$ will then be equivalent to the heat distribution of the same system where the energy, instead of being injected with a Gaussian distribution at time t' was injected in a point-like distribution at time t'-$\tau$.

It is then possible to estimate the temperature at the end of the bleaching cycle and its evolution (to calculate the temperature during the cycle more calculations are needed) over time. The temperature at the end of the bleaching cycle (whose timespan is $t_{bleaching}$) is the maximum temperature that the system experiences, therefore there is no need to assess the temperature during the bleaching cycle.

Then, with $P(t')$ the total power absorbed by the system, it is possible to write:

(j) $\Delta T(x,y,z,t) = \int_{-\tau}^{-\tau+t_{bleaching}} dt' \frac{P(t')}{\rho\, C_p} * \frac{1}{(4\pi\alpha)^{\frac{3}{2}}} *$

$\frac{1}{(t-t')^{\frac{3}{2}}} Exp\left(-\frac{((x-x')^2+(y-y')^2+(z-z')^2)}{4\alpha(t-t')}\right)$

If we consider the point at the center of the Gaussian (x=y=z=0) ant that the absorbed power is constant with the time, the integral is of easy solution:

(k) $\Delta T(x,y,z,t) = \int_{-\tau}^{-\tau+t_{bleaching}} \frac{dt'}{(t-t')^{\frac{3}{2}}} * \frac{P}{\rho\, C_p (4\pi\alpha)^{\frac{3}{2}}}$

$\Delta T(x,y,z,t) = \left[\frac{2}{\sqrt{t-t'}}\right]_{-\tau}^{-\tau+t_{bleaching}} * \frac{P}{\rho\, C_p(4\pi\alpha)^{\frac{3}{2}}}$

$\Delta T(x,y,z,t) = 2\left[\frac{1}{\sqrt{t-t_{bleaching}+\tau}} - \frac{1}{\sqrt{t+\tau}}\right] * \frac{P}{\rho\, C_p(4\pi\alpha)^{\frac{3}{2}}}$

Considering $t = t_{bleaching}$ the instant with the maximum temperature is:

(l) $\Delta T(x,y,z,t) = 2\left[\frac{1}{\sqrt{t+\tau}} - \frac{1}{\sqrt{t_{bleaching}+\tau}}\right] * \frac{P}{\rho\, C_p(4\pi\alpha)^{\frac{3}{2}}}$

substituting the numerical values. For the 1P case we have $P(100\%) = 0.88 * 10^{-3} W$ and $t_{bleaching} = 8 * 10^{-6}\, s$.

(m) $\Delta T(x,y,z,t) = 2\left[\frac{1}{\sqrt{1*10^{-7}\,s}} - \frac{1}{\sqrt{8.1*10^{-6}\,s}}\right] * \frac{0.88*10^{-3}W}{1.44*10^6\,J\bullet K^{-1}\bullet m^{-3}} * \frac{1}{(3.8*10^{-4}m^2\bullet s^{-1})^{\frac{3}{2}}}$

$\Delta T(x,y,z,t) = 2\left[3162\, s^{\frac{-1}{2}} - 351 s^{\frac{-1}{2}}\right] * 0.88 * 10^{-3}W * 9.25 * 10^{-2}(K\bullet s^{\frac{3}{2}} \bullet J^{-1})$

$\Delta T(x,y,z,t) = 2\left[2811\, s^{\frac{-1}{2}}\right] * 8.14 * 10^{-5}\left(K\bullet s^{\frac{1}{2}}\right) = 0.45\, K$

Therefore, the amount of heat released is negligible.

Actually, in the 1P case the geometry is slightly different from the one considered here. Heat diffusion occurs in only half (lower part – crystal corresponding) of the space. To take this into account is sufficient to consider that all the heat that diffused trough the surface is mirrored back into the perovskite crystal. The heat in the perovskite will then be doubled in respect of the current calculation bringing the temperature rise to 0.9 K.

For the 2P case, the calculation is also an approximation since the volume in which the energy is absorbed is actually larger, cylindrical like, than the one considered here. However, we do not expect qualitative differences in the temperatures since the considered volumes and energies are anyway similar.

We note that measuring transient temperature variations inside a material is a real technological challenge. While in principle possible(one can think of equipping the confocal microscope with means to sense the thermal-IR emission, during bleaching, but, as heat diffusion is relatively rapid, meaning that the signal would vary rapidly with the time), this needs detection with extremely high sensitivity; we do not know of existing or easily adaptable experimental setups that will allow such an evaluation.

To conclude, it has to be noted that the recombination dynamics is in the order of 5-10 ns, which is of the same order of the repetition time in the 2P experiment. Even if the 2P laser has pulses of 140 fs the heat release is mainly in the time-scale of the carrier recombination which, as we already mentioned, is in the same range of the pulse repetition rate. We can then neglect the decaying dynamics and consider the energy of the laser to be deployed uniformly in time while the laser is focused on the pixel with an error of few degrees.

We add as a final remark, some observation that can be a starting point for future work, that supports the validity of our calculation. Firstly, we did not find any difference of the 2P-PL at 25°C, after keeping the temperature of the single crystal at +100°C for 1 hour and then cooling back to +25°C. Thus, for damage due to temperature variations to affect the PL, the local temperature of the sample should be > 100°C. We also performed experiments, similar to the ones we report as 2P in Fig. 1, while the sample is at different temperatures and find that the higher the temperature (+50°C, +60°C, +80°C, +100°C) the lower the PL after bleaching. Thus, bleaching at 16% 2P-LP, in $MAPbBr_3$ leads to PL decrease as the temperature increases, instead of PL increase, as is the case at RT. If the bleaching increases the temperature significantly (more than 100°C) then an initial temperature difference before bleaching (passing from +25°C to +50°C) should not cause the photodamage to be qualitatively different. The temperature after bleaching should be, indeed, anyway elevated. The strong dependence of the bleaching effect on the initial temperature suggests that the photoinduced degradation happens at temperatures similar to the ones before bleaching.

## SI.7. Spectra and FLIM measurements.

The spectra and FLIM measurements were recorded with the same 2P confocal microscope used for the FRAP experiments. Spectra were recorded with the Zeiss 880 LSM detector with a 3 nm spectral resolution, using the same laser settings as those used in the FRAP experiments. With those settings the acquisition time of a single image is long (~5 min), because of the need to repeat the imaging for every wavelength. The crystals were exposed to different bleaching cycles than those used to determine the self-healing recovery and threshold of damage. In particular, the crystals were bleached 10 times to increase the time needed to heal the damage and avoid undesired artifacts. The spectra were taken successively. Repeating the data acquisition just after the first time did not give differences in the recorded spectra. For FLIM, a fast Big2 GaAsP detector was used. The pixel dwell time had to be increased to collect a sufficient number of photons to define a decay curve separated into 256 time-steps. A pixel dwell time of ~40 µs was used and the scans were repeated for a total of 16 times to reach the desired photon quantity for a total acquisition time of 120 sec/image.

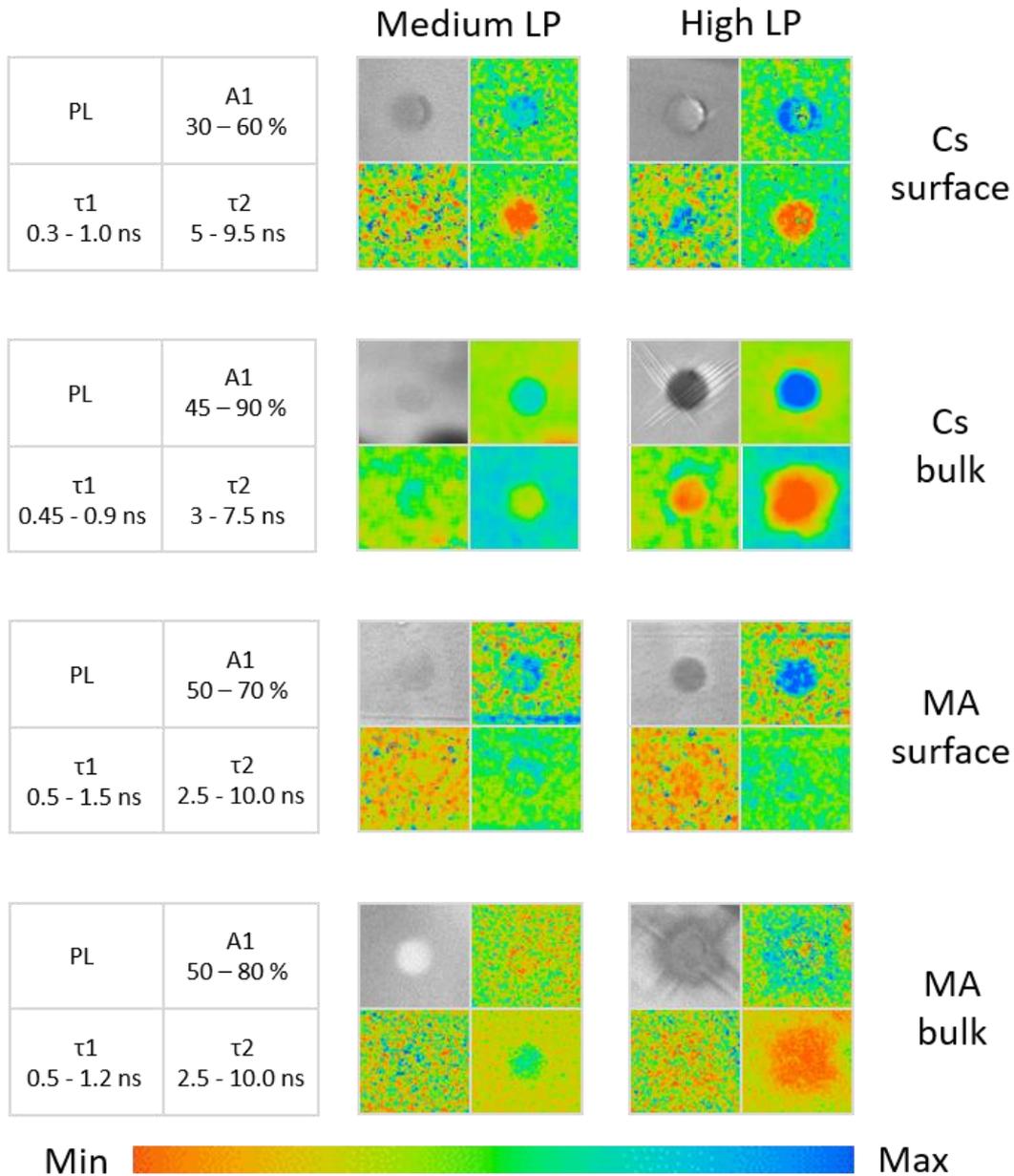

**Figure S3 :** Values of the parameters of the bi-exponential fits (equation *(g)*) of the PL decay for $CsPbBr_3$ and $MAPbBr_3$ crystals at their surfaces and in their bulk. Each square in the 4 squared images in the MIDDLE and RIGHT columns, is made up of the values of PL (upper left corner), A1 (upper right corner), τ1 (lower left corner) and τ2 (lower right corner) for the different settings and samples. The LEFT column then gives the minimum and maximum values for each corresponding square in the 4 squares **in the MIDDLE and RIGHT** columns. These correspond respectively to intermediate and high intensity bleaching. The bar at the bottom provides the color code for the images.

Thus, the images reported in Figures 2b, 2c in the main text were collected after bleaching the single crystals with 10 bleaching cycles. Laser powers of 6% and 14% were used to bleach the Cs sample on the surface, 15% and 22% for bleaching it in the bulk. For MA, laser powers of 10% and 14% were used to bleach the surface and 18% and 26% for bleaching in the bulk. The difference between the values of the bleaching laser power in the bulk and on the surface is due to the above-described aberration effect when the microscope is focused in the bulk.

The recorded decay curves were analyzed with the SPCImage software from Becker & Hickl GmbH. Because the repetition rate of the pulsed laser (12.5 ns) is in the same time scale as the decay, the program takes into account the incomplete decay of the PL, when a new pulse impacts the crystal again. A bi-exponential decay is used to fit the data:

$$(n) \; PL(t) = A1 * e^{\frac{-t}{\tau 1}} + B2 * e^{\frac{-t}{\tau 2}}$$

with $\tau_1$ and $\tau_2$ the shortest and longest decay times, $A1$ and $B2$ the prefactors for the respective fast and slow decays. This fitting is adequate to describe a simple model where the PL decay is determined by the trapping and trap-assisted recombination of the excited minority carriers. Under the conditions considered, the electron-hole recombination that leads to the PL is not the dominant recombination mechanism at any moment of time of the decay. Defining $k_{bi}$ as the kinetic constant of the bimolecular recombination, $n_M$ the majority carrier concentration, $k_{trap}$ the kinetic constant of the trapping of the minority carrier and $k_{recomb}$ the kinetic constant of the recombination between the majority carriers and the trapped minority carriers (see figure S6-left) this condition corresponds to

$$(o) \begin{cases} k_{bi} n_M \ll k_{recomb} \\ k_{bi} n_M \ll k_{trap} \end{cases}$$

Whether these inequalities hold or not is verified in two ways 1) the obtained fitting parameters should not vary with increasing/decreasing laser power, used to excite the PL for the FLIM experiment (which, if a bimolecular recombination mechanism were present, should modify $\tau_1$, because of the increasing density of both carriers with increasing absorbed radiation). 2) in the bleached volumes/areas, the obtained fitting parameters vary with respect to the backgrounds in ways that cannot be modeled by a bimolecular recombination decay. In Figure S3 we report, similarly to what is shown in Figures 2b and 2c of the main text, the PL and the lifetime ($\tau_2$) – corresponding to trap-assisted recombination – of the bleached areas for the Cs and MA samples both on the surface and in the bulk. This time however, we also report the shorter lifetime ($\tau_1$) – corresponding to trapping – as well as the value of the A1 parameter (cf. eqn. (i)) of the faster decay (A1 and B2 are normalized to 100% → B2=100%-A1). In the figure, each of these parameters is mapped with a color map. For each experiment the colors have different meaning and intervals which is reported in the left column. As is seen readily, A1 and $\tau_2$ vary substantially after bleaching (esp. for the Cs compound), while $\tau_1$ is not strongly modified. In a one trap model (Figure S4a) we can assume two different probabilities of trapping $k_{trap}$ and recombination through the trap $k_{recomb}$ (probability for the electron in the trap to recombine with a hole in the valence band, assuming electrons as minority carriers). In such a case, we can obtain the reported changes in the values of the fitting parameters when increasing the number of traps $N_t$ (or reducing them for the case which shows an increased luminescence). This is shown clearly in Figure S.4b where we simulate the values of A1, B2 $\tau_1$ and $\tau_2$ as function of the ratio between the trap concentration $N_t$ and the number of photo-excited electrons in the conduction band $n_c$ or holes in the valence band $h_v$ at t=0 of the FLIM excitation. These fitting numbers were obtained by fitting of the calculated PL obtained by the described model. In particular, the PL is proportional to $n_c * h_v$, the product of electron and hole densities. To obtain the value of this product we solved the following system of equations, valid for an intrinsic semiconductor with a single trap level:

$$(p) \begin{cases} \frac{dn_c}{dt} = -k_{trap} * n_c * h_t - k_{bi} n_c * h_v \\ \frac{dn_t}{dt} = k_{trap} * n_c * h_t - k_{recomb} * n_t * h_v \\ \frac{dh_v}{dt} = -k_{recomb} * n_t * h_v - k_{bi} n_c * h_v \\ N_t = n_t + h_t \\ n_c(t=0) = 1 \\ h_v(t=0) = 1 \end{cases}$$

with $n_t$ electrons and $h_t$ holes in the trap level and $k_{bi} = 0$. By normalizing to the value of $n_c * h_v$ at t = 0 and fitting the graph with equation (g) we obtained the values A1, B2, $\tau_1$ and $\tau_2$.

As previously described, a bi-exponential decay is obtained in a situation in which the trapping process is much faster than the trap-assisted recombination. This corresponds to the condition $k_{trap} \gg k_{recomb} \gg k_{bi}$ in the equations (*i*). If $k_{trap} < k_{recomb}$, a substantially mono-exponential, $k_{trap}$ determined, decay would be found. The parameters to obtain the evaluation of the fitted parameter are reported in Figure S4. Upon increasing the number of traps (when it is still lower than the number of excited carriers $\frac{n_t}{n_c(t=0)} < 1$) we see an increase of A1, a decrease of $\tau_2$ and little variation in $\tau_1$. This trend follows what we reported in Figure S3 for the Cs and Ma sample at high bleach intensity. In the case of MA of low bleaching intensity, the values behave in an opposite way, which, in this model, is interpreted as a decrease of the trap number.

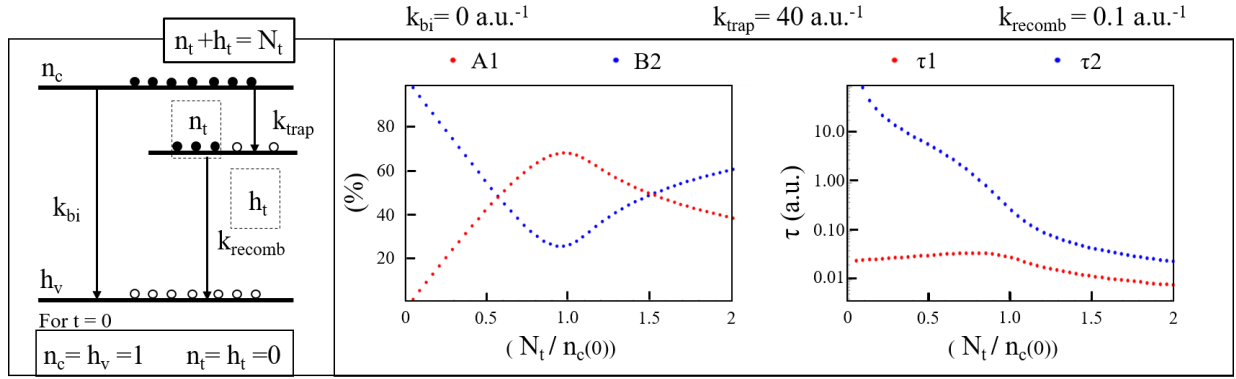

**Figure S4 :** (**left**) Scheme of the model used to derive equations in (i). (**center**) % of the first (A1) and second (B2) component of the fitting through a bi-exponential decay of the solution of equations (g). The two components are shown as function of the ratio between the number of traps ($N_t$) and the initial electron density, $n_c$. As can be seen, the component of the shortest lifetime (red) is the highest for a ratio approximately equal to 1. In our case, the results are consistent with an initial situation where the number of traps is smaller than the number of carriers. After the bleach, the number of traps increases (decreases for MA – medium power). Comparing with Figure.S3 is possible to see how the values of A1, B2, τ1 and τ2 follow the tendency described here for trap numbers for $N_t/n_c$<.1 increasing (decreasing for MA – medium power) after the bleach. (**right**) shortest (red) and longer (blue) decay time of the fitting through a bi-exponential decay of the solution of equations (i). The two times are shown as function of $N_t/n_c$. As can be seen the longest lifetime is the most sensitive to the variation of the number of traps while the shortest one does not vary much. This behavior is also consistent with the results reported in Figure.S3.

## SI.8. Computational Details and calculated structures.

All DFT calculations were performed using the Perdew-Burke-Ernzerhof (PBE)[8] form of the generalized-gradient approximation, augmented by dispersion terms calculated within the Tkatchenko-Scheffler (TS) scheme[9]. All calculations were performed using the Vienna ab initio simulation package (VASP)[10], a plane-wave basis code in which ionic cores are described by the projected augmented wave (PAW)[11] method. A plane-wave cutoff of 500 eV, 600 eV and 800 eV was used for the $CsPbBr_3$, $MAPbBr_3$ and $FAPbBr_3$ systems, respectively. A $10^{-6}$ eV/supercell convergence criterion for the total energy was used in all calculations.

3×3×3 supercells of the cubic $MAPbBr_3$ and $FAPbBr_3$ systems (a=5.92Å and a=5.97Å, respectively)[12] and 2×2×2 supercells of the orthorhombic $CsPbBr_3$ system (a=8.24Å, b=11.74Å, c=8.20Å)[13] were used in all calculations to mitigate defect-defect interactions. Consequently, a 2×2×2 k-point grid was used for all supercell calculations.

To introduce a MABr vacancy into the system a MA molecule and a nearby Br atom were taken out of the supercell. Similarly, in order to create an HBr vacancy, both MA and Br were taken out, and CH$_3$NH$_2$ was placed in the MA position. To create a CH$_3$NH$_2$ vacancy, MA was taken out of the supercell and an H atom was put in its place. To create a Br interstitial in all systems, a Br atom was added to the supercell in several different locations in order to determine its ideal, minimum energy position. All structures were then allowed to relax until the forces acting on the ions were below 0.015 eV/Å.

In the main text, we reported the structure for the positive charge interstitial defect of the CsPbBr$_3$, MAPbBr$_3$ and FAPbBr$_3$ systems. Defects can however be charged differently in function of the Fermi energy of the system. In Figure S5 we report the structure and electron density for the positive, neutral and negative Br interstitial defects in MAPbBr$_3$ ((a), (b) and (c) respectively) and CsPbBr$_3$ ((d), (e) and (f) respectively). We were not able to determine the equilibrium structures of the neutral and negative Br interstitial defects in FAPbBr$_3$ due to convergence issues. As can be seen the structure of the neutral and negative interstitial defects ((b) and (c), (e) and (f)) do not differ much. This suggests that the transition between the neutral and the negative state of the defect should be less energetically costly than the transition between the positively charged and the neutral one. The partial charge density associated with the defect is also not strongly affected by the added electron. In particular, no difference is easily identifiable for the Cs sample and any such difference is negligible for the MA sample.

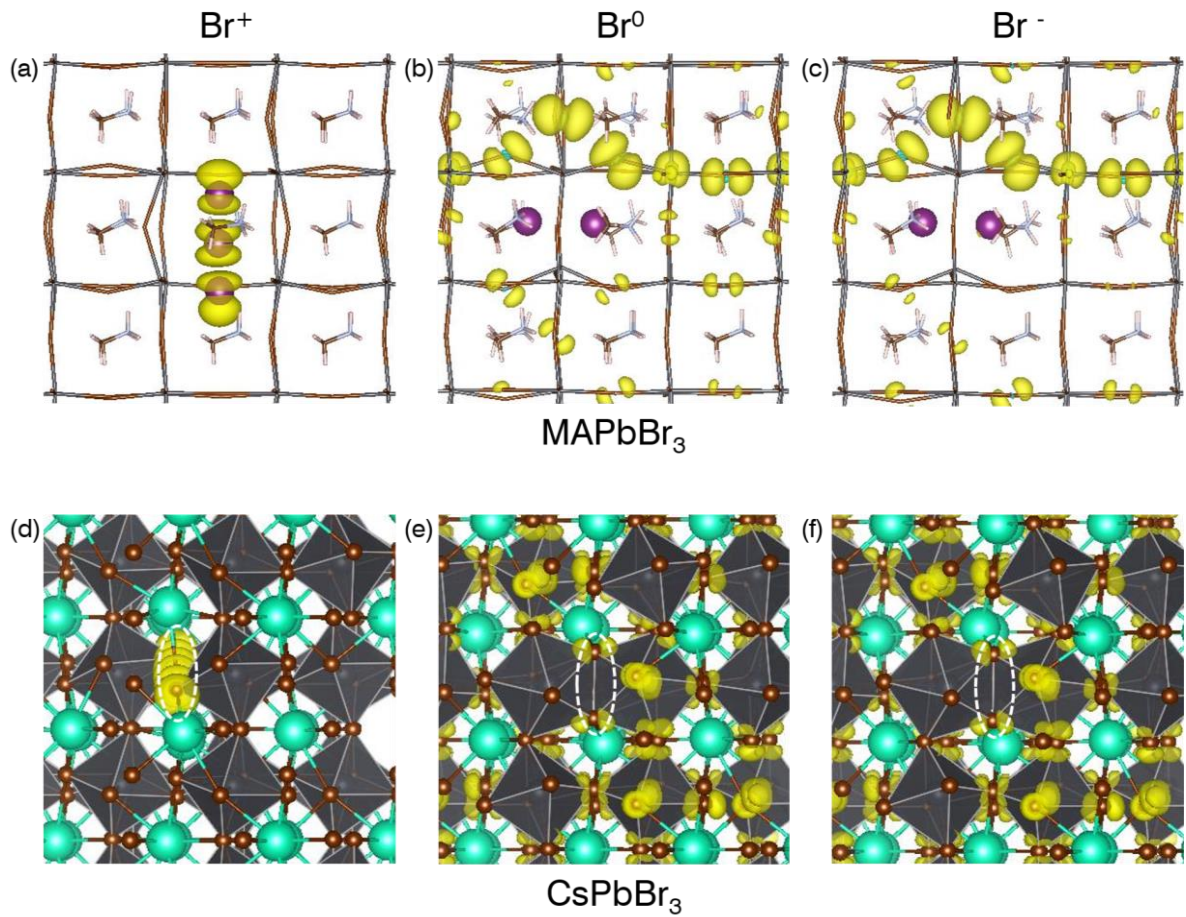

**Figure S5**: TOP: The minimum energy structure of MAPbBr$_3$ containing (a) Br$_i^+$ (b) Br$_i^0$ and (c) Br$_i^-$. The Br$_i$ defects are marked in purple, and the yellow contours represent the partial charge density associated with the defect's eigenvalue.
BOTTOM: The minimum energy structure of CsPbBr$_3$ containing (d) Br$_i^+$ (e) Br$_i^0$ and (f) Br$_i^-$. The Br$_i$ defects are marked by a dashed oval, and the yellow contours represent the partial charge density associated with the defect's eigenvalue.

## SI.9. Halo around the bleached area for the 2P–high laser power experiment

When the laser power switches from 0% to the desired % during the scanning of the bleaching cycle there is a short transient in which the laser power is not at the desired value. This is due to the Acousto-Optic-Modulator governing the laser power, which leads to an error over 1-2 pixels. While acceptable for most 2P confocal microscopy applications, in our case, the laser power during this transient is in the range that produces the state of higher luminescence, causing the "ring-effect" in Figure.1f-MA. Similar ring-effects were also found in other experiments with MA samples (not shown in this report).

## SI.10. AFM measurement

AFM measurements were carried out in a Smart AFM (Horiba); Scans of 1000×300 pixels per image were made in the AC mode at a 0.5 Hz scanning speed, using a highly-doped silicon probe (Olympus Co. AC160) with nominal resonance frequency and spring constant of 300 kHz and 26 N•m$^{-1}$. The sample was prepared by bleaching a long rectangular shape on the surface of a MAPbBr$_3$ single crystal. The bleaching was repeated 100 times in the 1P confocal microscope on the same area using 100% of the laser power. In this way, the halo around the bleached area reached farther from the bleached area itself. Attempts to use the AFM on the bleached area directly or close to the bleached area failed, probably because of the presence of the liquid-like phase, shown in Figure 4a of the main text. By measuring close to the bleached area, it was possible to obtain repeatable results. The distance between the bleached area and the right part of Figure 4a was estimated at 20 - 50 μm. An optical image of the bleached area (using the opticalmicroscope, connected to the AFM) and a scheme of the positioning of the AFM image with respect to the bleached area are shown in Figure S6.

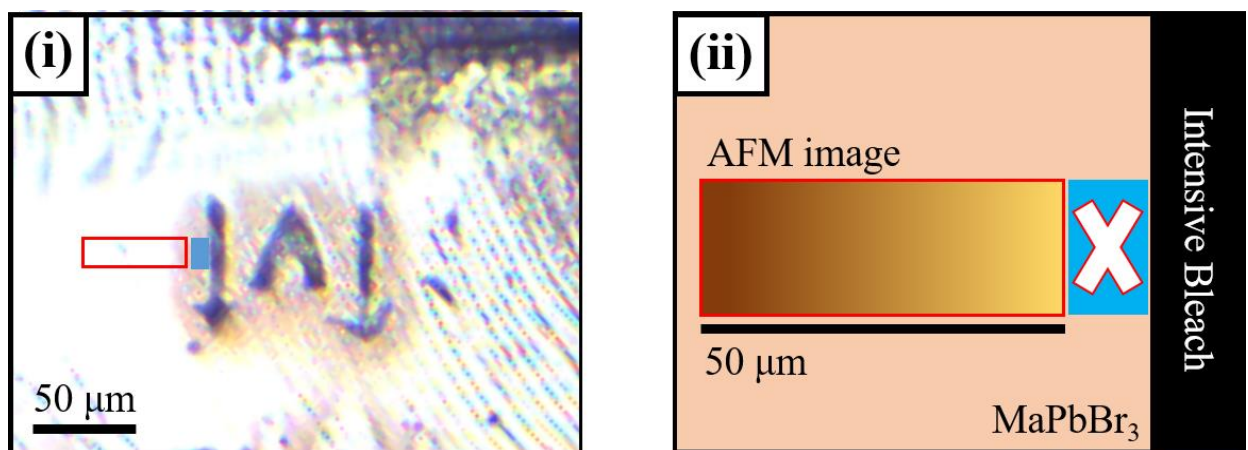

**Figure.S6**: (i) Optical image of the AFM scanned area. The direction of the arrows indicates the confocal scan direction. The letter A was used to indicate that in this bleaching spot the maximum laser intensity was used to perform the bleaching. Other experiments were performed with lower laser intensities but the results were less clear. (ii) Scheme of the position of the scanned area with respect to the bleached area.

## SI.11. Products of decomposition and formation of a liquid phase.

In the main text we described how the methylamine, produced by the 1-photon decomposition of $MAPbBr_3$, is probably the cause of the formation of $MAPbBr_3$ nanoparticles embedded in a liquid-like phase on the surface of the single crystal. Methylamine, $CH_3NH_2$, has been reported as one of the main decomposition products of $MaPbBr_3$, the others being $CH_3Br$, $NH_3$ and $HBr$ and $Br_2$. Exposing the perovskite to $CH_3Br$ and $HBr$ does not produce any visible effect (tested in the laboratory), while exposure to $CH_3NH_2$ and $NH_3$ causes the formation of a liquid phase as reported in the literature[14]. We already discussed in the main text how the lower boiling point of $NH_3$ implies that most of the material that re-condenses and forms the liquid phase has to be $CH_3NH_2$. $NH_3$ would indeed evaporate much quicker. Exposure to $Br_2$ leads, instead, to whitening of the single crystals (as we could show in separate experiments in our laboratory).

In the case of $FAPbBr_3$, the decomposition leads to $HBr$, $Br_2$, $NH_3$ and 1,3,5-triazine $(HCN)_3$ [15,16], which is an heterocyclic aromatic ring compound with melting point around 80°C[17]. As the first three molecules have already been analyzed, the triazine can also be excluded to form any liquid phase because of its high melting point and its acidity (liquids form with basic methylamine). We note that, even if formamidine can be synthesized by the reaction of formamidinium with a strong base[18], it has not been reported hitherto as decomposition product of halide perovskites. In the case of $CsPbBr_3$, the only gaseous decomposition product is $Br_2$, which does not form any liquid with the perovskites.

## SI.12. SEM and EDS measurements.

SEM and EDS studies were performed using a Zeiss Supra55 or Ultra55) field emission SEM equipped with Bruker XFlash 6, 60 mm detector. A landing voltage of 10 kV was applied on the samples, using a 30 μm aperture.

Quantification of the Pb:Br ratio requires calibration, because the e-beam affects different elements (and electronic transitions in an element) differently, leading to differences in X-ray emission. In particular, to obtain absolute atomic concentration ratios one has to have measured them from analytical chemistry tests on a reference material, which is similar to the sample. The EDS signal of each atom is multiplied by a calibration factor to give the chemically-measured atomic ratio in the reference sample. EDS can then be used to determine variations of the given ratio of elements in the sample of interest. In particular we have in expression (q) the relation between the real atomic ratio $\frac{Pb_{real}}{Br_{real}}$ and the EDS measured one $\frac{Pb_{EDS}}{Br_{EDS}}$.

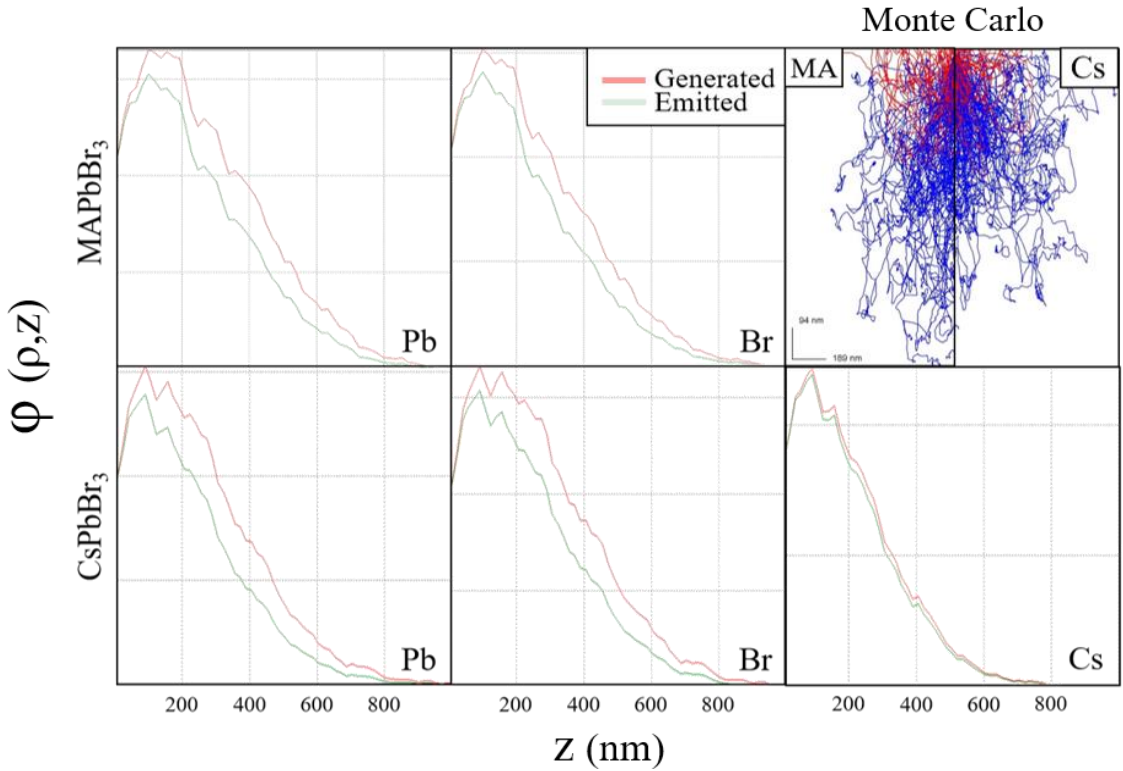

**Figure.S7** :.Phi-Rho-Z (φρz) curves for $MaPbBr_3$ (first line) and $CsPbBr_3$ (second line) as function of z, the depth inside the crystal. Since not all the generated X-rays are emitted from the surface we report two curves. The red curve corresponds to the amount of generated X-rays in function of z. The green curve is the amount of the X-rays effectively able to exit the surface in function of z. Curves for LEFT: Pb; MIDDLE: Br; RIGHT: Cs. In the upper right corner, we show graphically the simulation of the e-beam scattering into the $MaPbBr_3$ (left) and $CsPbBr_3$ (right).

$$(q) \quad \frac{Pb_{real}}{Br_{real}} = \frac{cal_{Pb}}{cal_{Br}} * \frac{Pb_{EDS}}{Br_{EDS}}$$

As both are known in the reference sample one can determine the value of $\frac{cal_{Pb}}{cal_{Br}}$ which can then be used and multiplied by $\frac{Pb_{EDS}}{Br_{EDS}}$ of the sample of interest to determine $\frac{Pb_{real}}{Br_{real}}$.

In our case, we were not interested in the Pb/Br ratio in the material, but in the variation of this ratio between the bleached and not-bleached zones. This can be easily determined without performing any calibration since the calibration factors cancel when taking the ratio as in (r)

$$(r) \quad \frac{\frac{Pb_{real}^{Bleach}}{Br_{real}^{Bleach}}}{\frac{Pb_{real}^{Ref}}{Br_{real}^{Ref}}} = \frac{\frac{cal_{Pb}}{cal_{Br}} * \frac{Pb_{EDS}^{Bleach}}{Br_{EDS}^{Bleach}}}{\frac{cal_{Pb}}{cal_{Br}} * \frac{Pb_{EDS}^{Ref}}{Br_{EDS}^{Ref}}} = \frac{\frac{Pb_{EDS}^{Bleach}}{Br_{EDS}^{Bleach}}}{\frac{Pb_{EDS}^{Ref}}{Br_{EDS}^{REf}}}$$

In the main text we referred to the fact that the EDS penetration depths actually create artifacts in evaluating the variation of the composition on the surface, because some signal comes from deeper layers. We evaluated this issue with the software Win X-Ray 1.4.2[19] for the incident voltage of 10 kV. As can be seen in Figure S7, the peak of the signal comes from around 150 nm. If we integrate the signal coming from the first 50 nm, we see that it corresponds to approximately 1/10 of the total integral. Figure S7 shows the results of the calculated Phi-Rho-Z ($\varphi\ \rho\ z$) curves for MaPbBr$_3$ (first line) and CsPbBr$_3$ (second line) as function of z, the depth inside the crystal. The value on the y axis, namely phi ($\varphi$), corresponds to the electron beam-induced ionization distribution in the material. In red is reported the origin and amount of the generated X-rays in the material. In green is the origin and amount of the X-rays effectively able to be emitted by the material, because part of them is reabsorbed. The amount of generated and emitted X-rays depends on the density of the material rho ($\rho$)(MA 3.83 g•cm$^{-3}$, Cs 5.4 g•cm$^{-3}$) and from the depth (z).

## SI.13. XPS measurements.

XPS measurements were carried out with Kratos AXIS ULTRA system using a monochromatic Al K$\alpha$ X-ray source (h$\nu$ = 1486.6 eV) at 75W and detection pass energies ranging between 20 and 80 eV. Curve fitting analysis was based on linear or Shirley background subtraction and application of Gaussian-Lorenzian line shapes.

## SI.14. Bibliography